\documentclass[12pt]{article}
\usepackage{times}

\usepackage{epsf}
\def\beq{\begin{equation}}
\def\eeq{\end{equation}}
\def\bea{\begin{eqnarray}}
\def\eea{\end{eqnarray}}

\def\nnb{\nonumber}

\def\rar{\rightarrow}
\def\nnb{\nonumber}

\def\ba{\begin{array}}
\def\ea{\end{array}}
\def\bea{\begin{eqnarray}}
\def\eea{\end{eqnarray}}
\def\BKsll{$B\rightarrow \,K^*\, \ell^+ \ell^-$}
\def\BKll{$B\rightarrow \,K\,\ell^+ \ell^-$}
\def\BKstt{$B\rightarrow \,K^*\, \tau^+ \tau^-$}
\def\BKtt{$B\rightarrow \,K\, \tau^+ \tau^-$}
\def\tep{$b \rar s \, \ell^+ \ell^-$}

\title{The exclusive $B\rightarrow (K,K^*)\ell^+\ell^-$ decays in a CP softly
broken  two Higgs doublet model}
\author{\vspace{1cm}\\
         {\bf G\"{u}ray Erkol}
         \thanks{E-mail address:
        gurerk@newton.physics.metu.edu.tr} \, \, and \, \,
        {\bf G\"{u}rsevil  Turan}
        \thanks{E-mail address:
        gsevgur@metu.edu.tr}\,\,\thanks{Common address: Middle East Technical University Physics Dept. Inonu Bul.
06531 Ankara-TURKEY}}
        \date{}
\begin{document}
\setlength{\baselineskip}{24pt} \maketitle
\setlength{\baselineskip}{7mm}

\abstract{We study the differential branching ratio, forward-backward asymmetry,
CP-violating asymmetry, CP-violating asymmetry
in the forward-backward asymmetry and polarization asymmetries in the \BKll and \BKsll 
decays in the context of  a CP 
softly broken two Higgs doublet model. We analyze
the dependencies  of these observables on the model parameters by paying a special
attention to the effects of neutral Higgs boson (NHB) exchanges and possible CP violating
effects. We find that NHB effects are quite significant for both decays. A combined analysis of
above-mentioned observables seems to be very promising
as  a testing ground for new physics beyond the SM, especially for the existence  
of the CP-violating phase in the theory.

\thispagestyle{empty} \setcounter{page}{1}
\section{Introduction \label{s1}}
At the quark level, \BKsll and \BKll decays ($\ell=e,\mu ,\tau$) are 
induced by the  \tep transition, which has received considerable attention
\cite{gsw}-\cite{ks}, as a potential testing ground for the effective Hamiltonian describing 
the flavor changing neutral current processes in B decays. They are also expected
to open a window to investigate the new physics prior to any possible experimental
clue about it.

It is well known that the inclusive rare decays, although theoretically cleaner 
than the exclusive ones, are   more   difficult to measure. This fact stimulates the study
of the exclusive decays, but the 
situation is contrary then: their experimental study is easy but the theoretical
investigation is hard. For inclusive semileptonic B-meson decays, the physical 
observables can be calculated in heavy quark effective theory (HQET) \cite{HQET};
however the description of the  exclusive decays requires the additional knowledge
of decay form factors, i.e., the matrix elements of the effective Hamiltonian
between the initial B and final meson states. Finding these hadronic transition
matrix elements  is related to the nonperturbative sector of the QCD and should be 
calculated by means of a nonperturbative approach. The form factors for B decays
into $K$ and $K^*$ have been calculated in the framework of different 
methods, such as chiral theory \cite{CT}, three point QCD sum rules method \cite{QCDSR},
relativistic quark model \cite{Jaus}, effective heavy quark theory \cite{Roberts}, and
light cone sum rules \cite{Aliev1}, \cite{ PBALL}. 

From the experimental side,  there  exist  upper limits on the branching ratios
of $B^0\rightarrow K^{0*} \mu^+ \mu^-$ and $B^+\rightarrow K^+ \mu^+ \mu^-$, given by
CDF collaboration \cite{Affolder}
\bea
BR (B^0\rightarrow K^{0*} \mu^+ \mu^-) & < & 4.0 \times 10^{-6} \nnb \\
BR (B^+\rightarrow K^+ \mu^+ \mu^-) & < & 5.2 \times 10^{-6} \nnb .
\eea
With these measured upper limits and  also the  recent measurement of the branching ratio of 
$B\rightarrow K \ell^+ \ell^-$ with $\ell=e,\mu$, 
\bea
BR (B\rightarrow K \ell^+ \ell^-) & = &( 0.75^{+0.25}_{-0.21}\pm 0.09) \times 10^{-6} \nnb ,
\eea
at KEK \cite{Belle}, the processes $B\rightarrow (K,K^*)\ell^+\ell^-$ have received great 
interest so that their theoretical calculation  has been the subject of many 
investigations in the SM and beyond, such as  the SM with 
fourth generation, multi-Higgs doublet models, minimal supersymmetric extension of the 
SM (MSSM) and in a model independent method \cite{dt}-\cite{erhan}. 

In this paper we will investigate the 
exclusive  $B\rightarrow (K,K^*)\ell^+\ell^-$ decays  in a CP softly broken  
 two Higgs doublet model, which is called model IV
in the literature \cite{Huang}.

CP violating asymmetry $A_{CP}$ is an important observable that may provide 
valuable information about the models used. In the SM the source of CP violation
is the complex  Cabibbo-Kobayashi-Maskawa 
(CKM) matrix elements and due to  unitarity of this matrix together with the smallness 
of the term $V_{ub} V^*_{us}$, $A_{CP}$ for $B\rightarrow (K,K^*)\ell^+\ell^-$ decays
almost vanishes in the SM. However, 
like many extensions of the SM, model IV predicts a  new source of CP violation so that
we have an opportunity to investigate the physics beyond the SM by analysing the CP 
violating effects. 

In model IV, up-type quarks get masses from Yukawa couplings to the one Higgs doublet
$H_2$, and down-type quarks and leptons get masses from another Higgs doublet $H_1$.
In such  a  2HDM, all the parameters in the Higgs potential 
are real so that it is CP-conserving, but one allows  the real and imaginary
parts of $\phi^+_1\phi_2$ to have different self-couplings so that the phase $\xi$,
which comes from the expectation value of Higgs field,  can not be rotated away, 
which breaks the CP symmetry (for details, see ref \cite{Huang}). In model IV,
interaction vertices of the Higgs bosons and the down-type quarks and leptons depend on
the CP violating phase $\xi$  and the ratio $\tan \beta =v_2/v_1$, where
$v_1$ and $v_2$ are the vacuum expectation values of the first and the second Higgs doublet
respectively, and they are free parameters in the model.  The constraints on $\tan \beta $
are usually obtained from $B-\bar{B}$, $K-\bar{K}$ mixing, $b\rightarrow s \, \gamma$
decay width, semileptonic decay $b\rightarrow c \, \tau \bar{\nu}$ and is given by
\cite{ALEP}
\bea
0.7 \leq \tan \beta \leq 0.52 (\frac{m_{H^{\pm}}}{1 ~ GeV}) \, ,
\eea
and the lower bound $m_{H^{\pm}} \geq 200$ GeV has also been given in \cite{ALEP}.

In addition to the CP asymmetry $A_{CP}$, differential or total 
branching ratios and the forward-backward
asymmetries, polarization asymmetries  are also thought to play an important role in 
further investigations of
the structure of the SM and for establising new physics beyond it. 
It has been pointed out in refs.\cite{Hewett} and \cite{GK} that the longitudinal polarization 
$P_L$ of the final lepton  may be accessible in the 
$B\rightarrow (K,K^*)  \tau^+\tau^- $ mode in the near future. It has been shown  \cite{ks}
 that together with 
$P_L$, the other two orthogonal components of polarization, $P_T$ and $P_N$, 
are crucial for the $\tau^+\tau^-$ mode since these three components contain the 
independent, but complementary information because they involve different combinations 
of Wilson coefficients in addition to the fact that they are proportional to 
$m_{\ell}/m_b$. Lepton polarizations in $B\rightarrow K (K^*) \ell^+\ell^-$ decays are analyzed in the model II 
version of the  2HDM and in a general model independent way in refs.\cite{Aliev2}
(\cite{Aliev6}) and  
\cite{Aliev9} (\cite{Aliev8}), respectively. Ref.\cite{Yan} gives an analysis of the
lepton polarization asymmetries in the processes $B\rightarrow (K ,K^*) \ell^+\ell^-$
in a supersymmetric context.

As pointed out  before, (see f.eg.\cite{Dai}-\cite{Xiong}), in models with two Higgs 
doublets, like MSSM, 2HDM, etc., neutral Higgs 
boson (NHB) effects could contribute largely to the semileptonic rare B meson decays,
especially  for heavy
lepton modes and for large $\tan \beta$. However, in the literature there was a 
disagreement about the results of 
NHB exchange  diagrams  contributing to \tep transition in the context 
of the 2HDM \cite{Logan,Kruger3}. This situation  seems to be resolved now 
\cite{Kruger3,Huang2}, and in view of  new 
forms of the Wilson coefficients $C_{Q_1}$ and $C_{Q_1}$  due to NHB effects, it is 
quite worthwhile to return to    the exclusive processes 
$B\rightarrow (K,K^*) \tau^+\tau^-$ in order to investigate the NHB effects 
together with the CP violating effects in model IV.

The paper is organized as follows: In Sec. \ref{s2}, after  we give the effective Hamiltonian 
and the  definitions of the form factors, we introduce basic formulas of observables.
Sec. \ref{s3} is devoted to the numerical analysis and discussion of our results.

\section{Effective Hamiltonian and form factors \label{s2}}
At the quark level, the effective Hamiltonian describing the rare semileptonic 
\tep transition can be obtained by integrating out the top quark, Higgs bosons and
$W^{\pm}$, $Z$ bosons:
\begin{equation}
{\cal H}_{eff}  =  \frac{4G_{F}}{\sqrt{2}} V_{tb} V_{ts}^{*} (\sum_{i=1}^{10}
    C_{i}(\mu) O_{i}(\mu) + \sum_{i=1}^{10} C_{Q_i}(\mu) Q_i(\mu))\; ,
        \label{eq:Ham}
\end{equation}
where $O_{i}$ are current-current $(i=1,2)$, penguin $(i=1,..,6)$, 
magnetic penguin $(i=7,8)$ and semileptonic $(i=9,10)$ operators and $C_{i}(\mu )$ 
 are the corresponding Wilson coefficients renormalized at 
the scale $\mu$ \cite{Grinstein, Misiak}. The 
additional operators $Q_{i}, (i=1,..,10)$ and their Wilson coefficients are due to the 
NHB exchange diagrams, which can be found in \cite{Huang3,Kruger3,Huang2}.

Neglecting the mass of the $s$ quark, the above
Hamiltonian leads to the following  matrix element:
\begin{eqnarray}\label{genmatrix}
{\cal M} &=&\frac{G_{F}\alpha}{2\sqrt{2}\pi }V_{tb}V_{ts}^{\ast }%
\Bigg\{C_{9}^{eff}~\bar{s}\gamma _{\mu }(1-\gamma _{5})b\,\bar{\ell}%
\gamma ^{\mu }\ell +C_{10}~\bar{s}\gamma _{\mu }(1-\gamma _{5})b\,\bar{%
\ell}\gamma ^{\mu }\gamma _{5}\ell  \nonumber \\
&-&2C_{7}^{eff}~\frac{m_{b}}{q^{2}}\bar{s}i\sigma _{\mu \nu
}q^{\nu }(1+\gamma _{5})b\,\bar{\ell}\gamma ^{\mu }\ell \Bigg\},\nonumber\\
\end{eqnarray}
where $q$ is the momentum transfer. Here, Wilson coefficient $C_9^{eff}(\mu)$
contains a perturbative part and a part coming
from long-distance effects due to conversion of the real $\bar{c}c$ into
lepton pair $\ell^+ \ell^-$:
\begin{eqnarray}
C_9^{eff}(\mu)=C_9^{pert}(\mu)+ Y_{reson}(s)\,\, ,
\label{C9efftot}
\end{eqnarray}
where
\begin{eqnarray}
C_9^{pert}(\mu)&=& C_9^{2HDM}(\mu) \nonumber
\\ &+& h(z,  s) [ 3 C_1(\mu) + C_2(\mu) + 3 C_3(\mu) +
C_4(\mu) + 3 C_5(\mu) + C_6(\mu)] \nonumber \\&-&   \frac{1}{2}
h(1, s) \left( 4 C_3(\mu) + 4 C_4(\mu)
+ 3 C_5(\mu) + C_6(\mu) \right)\nnb \\
&- &  \frac{1}{2} h(0,  s) \left[ C_3(\mu) + 3 C_4(\mu) \right]
\\&+& \frac{2}{9} \left( 3 C_3(\mu) + C_4(\mu) + 3 C_5(\mu) +
C_6(\mu) \right) \nonumber \,\, ,
\end{eqnarray}
and $z=m_c/m_b$. The functions $h(z,s)$ arises from the one loop contributions
of the four quark operators $O_1$,...,$O_6$ and their explicit forms can be found in
\cite{Misiak}.
It is possible to parametrize  the resonance $\bar{c}c$ contribution
$Y_{reson}(s)$ in Eq.(\ref{C9efftot}) using a
Breit-Wigner shape with normalizations fixed by data which is given by \cite{AAli2}
\begin{eqnarray}
Y_{reson}(s)&=&-\frac{3}{\alpha^2_{em}}\kappa \sum_{V_i=\psi_i}
\frac{\pi \Gamma(V_i\rightarrow \ell^+
\ell^-)m_{V_i}}{s m^2_B-m_{V_i}+i m_{V_i}
\Gamma_{V_i}} \nonumber \\
&\times & [ (3 C_1(\mu) + C_2(\mu) + 3 C_3(\mu) + C_4(\mu) + 3
C_5(\mu) + C_6(\mu))]\, .
 \label{Yresx}
\end{eqnarray}
The phenomenological parameter $\kappa$
in Eq. (\ref{Yresx}) is taken as $2.3$ so as to reproduce the correct value
of the branching ratio $BR(B\rightarrow J/\psi ~ X\rightarrow X\ell\bar{\ell})=
BR(B\rightarrow J/\psi ~ X) BR(J/\psi\rightarrow  ~ X \ell\bar{\ell})$.

Next we proceed to calculate the  differential branching ratio $dBR/ds$, 
forward-backward asymmetry $A_{FB}$, CP violating
asymmetry $A_{CP}$, CP asymmetry in the forward-backward asymmetry $A_{CP}(A_{FB})$
and finally the lepton polarization asymmetries    of the \BKll and \BKsll
decays. In order to find these physically measurable quantities at
hadronic level, the necessary matrix elements are
$<M(p_M)|\bar{s}\gamma_\mu(1-\gamma_5)b|B(p_B)>$,
$<M(p_M)|\bar{s}i\sigma_{\mu\nu}q_\nu(1+\gamma_5)b|B(p_B)>$ and
$<M(p_M)|\bar{s}(1+\gamma_5)b|B(p_B)>$ for $M=K,~K^\ast$, which can be parametrized in terms
of form factors. Using
the parametrization of the form factors as in \cite{Aliev2} and \cite{abhh},
we find the amplitudes governing the \BKll and the \BKsll decays as follows:
\bea\label{matrixp} \cal{M}^{B\rightarrow K
}&=&\frac{G_F\alpha}{2\sqrt{2}\pi}V_{tb}V_{ts}^\ast\Bigg \{[2A_1
p_K^\mu+B_1 q^\mu]\bar{\ell}\gamma_\mu \ell+[2G_1 p^\mu +D_1
q^\mu]\bar{\ell}\gamma_\mu\gamma_5\ell
+E_1\bar{\ell}\ell+F_1 \bar{\ell}\gamma_5 \ell \Bigg\}~,\nnb\\
\eea
and
\bea \label{matrixBKsll}\cal{M}^{B\rightarrow K^\ast}&=&\frac{G_F
\alpha}{2\sqrt{2}\pi}V_{tb}V_{ts}^\ast\Bigg \{
\bar{\ell}\gamma_\mu\ell[2A\epsilon_{\mu\nu\lambda\sigma}
\varepsilon^{\ast\nu} p_{K^*}^\lambda p_B^\sigma +i B
\varepsilon^\ast_{\mu}-i C(p_B+p_{K^\ast})_\mu (\varepsilon^\ast
q)-i D
(\varepsilon^\ast q)q_\mu]\nnb\\
&+& \bar{\ell}\gamma_\mu \gamma_5 \ell[2E
\epsilon_{\mu\nu\lambda\sigma}\varepsilon^{\ast\nu}
p_{K^\ast}^\lambda
 p_B^\sigma +i F \varepsilon^\ast_{\mu} -i G(\varepsilon^\ast q)(p_B+p_{K^\ast})
-i H(\varepsilon^\ast q) q_\mu]+i \bar{\ell}\ell
Q(\varepsilon^\ast q)\nnb\\&+&i \bar{\ell}\gamma_5 \ell N
(\varepsilon^\ast q)\Bigg \} \eea 
where
\bea
A_1&=&C^{eff}_9 f^+ -2m_B C^{eff}_7 \frac{f_T}{m_B+m_K},\nnb\\
B_1&=&C^{eff}_9(f^+ +f^-)+2 C^{eff}_7
\frac{m_B}{q^2}f_T\frac{(m_B^2-m^2-q^2)}{m_B+m_K},\nnb\\
G_1&=&C_{10}f^+,\nnb\\
D_1&=&C_{10}(f^+ +f^-),\nnb\\
E_1&=&C_{Q_1}\frac{1}{m_b}[(m_B^2-m_K^2)f^+ +f^- q^2],\nnb\\
F_1&=&C_{Q_2}\frac{1}{m_b}[(m_B^2-m_K^2)f^+ +f^- q^2],\nnb\\
A&=&C^{eff}_9\frac{V}{m_B+m_{K^\ast}}+4\frac{m_b}{q^2}C^{eff}_7 T_1,\nnb\\
B&=&(m_B+m_{K^\ast})\Bigg( C^{eff}_9 A_1+\frac{4
m_b}{q^2}(m_B-m_{K^\ast})C^{eff}_7
T_2\Bigg),\nnb\\
C&=&C^{eff}_9\frac{A_2}{m_B+m_{K^\ast}}+
4\frac{m_b}{q^2}C^{eff}_7\Bigg(T_2+\frac{q^2}{m_B^2-m_{K^\ast}^2}T_3\Bigg),\nnb\\
D&=&2C^{eff}_9\frac{m_{K^\ast}}{q^2}(A_3-A_0)-4C^{eff}_7\frac{m_b}{q^2} T_3,\nnb\\
E&=&C_{10} \frac{V}{m_B+m_{K^\ast}}, \label{eqff}\\
F&=&C_{10}(m_B+m_{K^\ast})A_1,\nnb\\
G&=&C_{10}\frac{A_2}{m_B+m_{K^\ast}},\nnb\\
H&=&2C_{10}\frac{m_{K^\ast}}{q^2}(A_3-A_0),\nnb\\
Q&=&2C_{Q_1}\frac{m_{K^\ast}}{m_b}A_0,\nnb\\
N&=&2C_{Q_2}\frac{m_{K^\ast}}{m_b}A_0.\nnb 
\eea 
Here $f^+$, $f^-$ and  $f_T$ and $A_0$, $A_1$, $A_2$, $A_3$, $V$, $T_1$, $T_2$ and $T_3$
are the relevant form factors in $B\rightarrow K$ and $B\rightarrow K^*$ transitions,
respectively. For $B\rightarrow K$, we use the results calculated in the light cone 
QCD sum rules framework, which can be written in the following pole forms \cite{Aliev2}
\bea
f^+ (q^2) &=& \frac{0.29}{\left(1 - \displaystyle{\frac{q^2}{23.7}}\right)}~, \nnb \\
f^-(q^2) &=& - \frac{0.21}{\left(1 - \displaystyle{\frac{q^2}{24.3}}\right)}~, \nnb \\
f_T(q^2) &=& - \frac{0.31}{\left(1 - \displaystyle{\frac{q^2}{23}}\right)}~,\label{pfm}
\eea
As for the  $B\rightarrow K^*$ transition, we use the result of \cite{PBALL}, where
$q^2$ dependence of the form factors can be represented in terms of three parameters 
as given by
\bea
F(q^2) = \frac{F(0)}{1-a_F\,\frac{q^2}{m_B^2} + b_F \left
    ( \frac{q^2}{m_B^2} \right)^2}~, \nnb
\eea
where the values of parameters $F(0)$, $a_F$ and $b_F$ for the
$B \rar K^*$ decay are listed in Table 1. The form factors $A_0$ and $A_3$ in Eq. 
(\ref{eqff}) can be found from the following parametrization,
\bea\label{paramet} A_0&=&A_3-\frac{T_3~q^2}{m_{K^*} m_b},\nonumber\\
A_3&=&\frac{m_B+m_{K^*}}{2m_{K^*}}A_1-\frac{m_B-m_{K^*}}{2m_{K^*}}A_2.
\eea

\begin{table}[h]                    
\renewcommand{\arraystretch}{1.5}                        
\addtolength{\arraycolsep}{3pt}
$$
\begin{array}{|l|ccc|}
\hline
& F(0) & a_F & b_F \\ \hline
A_1^{B \rar K^*} &
\phantom{-}0.34 \pm 0.05 & 0.60 & -0.023 \\
A_2^{B \rar K^*} &
\phantom{-}0.28 \pm 0.04 & 1.18 & \phantom{-}0.281\\
V^{B \rar K^*} &
 \phantom{-}0.46 \pm 0.07 & 1.55 & \phantom{-}0.575\\
T_1^{B \rar K^*} &
  \phantom{-}0.19 \pm 0.03 & 1.59 & \phantom{-}0.615\\
T_2^{B \rar K^*} & 
 \phantom{-}0.19 \pm 0.03 & 0.49 & -0.241\\
T_3^{B \rar K^*} & 
 \phantom{-}0.13 \pm 0.02 & 1.20 & \phantom{-}0.098\\ \hline
\end{array}   
$$
\caption{$B\rar K^*$ transition form factors in ligt cone QCD sum rules .}
\renewcommand{\arraystretch}{1}
\addtolength{\arraycolsep}{-3pt}
\end{table}

Using Eqs. (\ref{matrixp}) and (\ref{matrixBKsll}) and
performing summation over final  lepton polarization, we get for the double
differential decay rates:

\begin{eqnarray}\label{ddrp}
\frac{d^2\Gamma^{B\rightarrow K}}{ds ~dz}&=&\frac{G_F^2
\alpha^2}{2^{11}\pi^5}|V_{tb}V_{ts}^\ast|^2 m_B^3
\sqrt{\lambda}~v~ \Bigg\{m_B^2 \lambda(1-z^2 v^2)|A_1|^2+ s(v^2
|E_1|^2+ |F_1|^2) \nonumber
\\&+& (m_B^2 \lambda(1-z^2 v^2) +16~r~m^2_{\ell})~|G_1|^2
+4~s~m_{\ell}^2~|D_1|^2\nonumber\\
&+&
4~m_{\ell}^2~(1-r-s)~Re[G_1D_1^\ast]+2~v~m_{\ell}~\sqrt{\lambda}~z~
Re[A_1E_1^\ast]\nonumber\\
&+&2~m_{\ell}~((1-r-s)~Re[G_1F_1^\ast]+s Re[D_1F_1^\ast]) \Bigg\}
\, ,\end{eqnarray}
and
\bea\label{ddrate} \frac{d^2\Gamma^{B\rightarrow
K^\ast}}{ds~dz}&=&\frac{\alpha^2 G_F^2}{2^{15} m_B \pi^5}|V_{tb}
V^*_{ts}|^2 \sqrt{\lambda_{\ast}}~v~\Bigg \{ 4~s~\lambda_{\ast}
(2+v^2(z^2-1)) |A|^2\nonumber\\ &+&4\, v^2 \,s\, m_B^4
\lambda_{\ast}(1+z^2) |E|^2 +16\, m_B^2\, s \,v\,
z\sqrt{\lambda_{\ast}}\Big(Re[B E^\ast]+Re[A F^\ast]\Big)
 \nonumber\\&+& \frac{1}{r}\Bigg[[\lambda_{\ast} (1-z^2 v^2)+2\, r_{\ast} s (5-2 v^2)]|B|^2
+m_B^4\lambda_{\ast}^2 (1-z^2 v^2)|C|^2\nonumber\\&+&
[\lambda_{\ast} (1-z^2 v^2)-2\, r_{\ast} s (1-4 v^2)]|F|^2
+m_B^4\lambda_{\ast}[(-1+r_{\ast})^2(1-v^2) z^2\nonumber\\&+&
(-1+z^2)(s t^2-8(1+r_{\ast})t^2-\lambda_{\ast})]|G|^2 +2\,m_B^2
\lambda_{\ast} W_{\ast}(1-z^2 v^2) Re[B C^\ast]\nonumber\\&-& 2\,
m_B^2 \lambda_{\ast} [W_{\ast} (1-z^2 v^2)-4t^2]Re[FG^\ast] +m_B^2
\lambda_{\ast}\Big(4 s m_{\ell}(m_{\ell}|H|^2+Re[H
N^\ast])\nonumber\\&+&s(|N|^2+v^2 |Q|^2) -4 t (Re[F(2 t
H^\ast+N^\ast/m_B)]\nonumber\\ &+&4 (1-r_{{\ast}})m_{\ell} Re[G(2
m_{\ell}H^\ast+N^\ast)]\Big)\nonumber\\ &+& 4 t m_B v z^2
Re[(W_{{\ast}} B+m^2_B (W^2_{{\ast}}-4 r_{{\ast}}s)
C)Q^\ast]\Bigg]\Bigg\}\, .  \eea
Here $s=q^2/m_B^2$, $r_{(\ast)}=m_{K(K^\ast)}^2/m_B^2$,
$v=\sqrt{1-\frac{4t^2}{s}}$, $t=m_l/m_B$,
$\lambda_{(\ast)}=r_{(\ast)}^2+(s-1)^2-2r_{(\ast)}(s+1)$, $W_{(*)}=-1+r_{(*)}+s$ and
$z=\cos\theta$, where $\theta$ is the angle between the
three-momentum of the $\ell^-$ lepton and that of the B-meson in
the center of mass frame of the dileptons $\ell^+\ell^-$.

Having established the double differential decay rates, let us now consider the 
forward-backward asymmetry  $A_{FB}$ of the lepton pair, which is defined as 
\begin{eqnarray}
A_{FB}(s)& = & \frac{ \int^{1}_{0}dz \frac{d^2 \Gamma }{ds dz} -
\int^{0}_{-1}dz \frac{d^2 \Gamma }{ds dz}}{\int^{1}_{0}dz
\frac{d^2 \Gamma }{ds dz}+ \int^{0}_{-1}dz \frac{d^2 \Gamma }{ds
dz}}~~. \label{AFB1}
\end{eqnarray}
The $A_{FB}$'s for the \BKll and \BKsll  decays are calculated to be
\bea A_{FB}^{B\rightarrow K}&=&-\int\,\,ds\,\, (t v^2
\lambda Re(A_1~E_1^\ast))\Bigg/ \int\,\,ds\,\, v
\sqrt{\lambda} \, \Delta ~,\label{piafb} \eea
\bea\label{afbrho}
A_{FB}^{B\rightarrow K^\ast}&=&\int~ds~2m_B^3\lambda_\ast v^2 \Big( 4m_Bs(Re[B~E^\ast]+Re[A~F^\ast])\nnb\\
&+&\frac{t}{r_{\ast}}[W_\ast Re[B~Q^\ast]+m_B^2\lambda_\ast
Re[C~Q^\ast]]\Big)\Bigg/\int~ds \sqrt{\lambda_\ast} ~~v~~
\Delta_{\ast}.
 \eea

We note that in the SM, $A_{FB}$ in \BKll decay is zero because of the fact that 
hadronic current for 
$B\rar K$  transition does not have any axial counterpart. As seen from 
Eq.(\ref{piafb}), it is also zero in model IV unless we do not take into effect 
the NHB exchanges. Therefore,  \BKll decay may be a good candidate for testing the 
existence and the importance of NHB effects.

In this work, we also analyse the CP violating asymmetry $A_{CP}$, which is defined as
\begin{eqnarray}
A_{CP}& = & \frac{d\Gamma/ds (B\rar M \, \ell^+ \ell^-)
-d\Gamma/ds (\bar{B}\rar \bar{M} \, \ell^+ \ell^-) }{d\Gamma/ds (B\rar M
\, \ell^+ \ell^-)  + d\Gamma/ds (\bar{B}\rar \bar{M} \, \ell^+ \ell^-)}
~~. \label{ACP1}
\end{eqnarray}
where  $M=K,~K^\ast$ and $d\Gamma/ds$ are the corresponding differential decay rates, which
are obtained by integrating the expressions in Eqs. (\ref{ddrp}) and 
(\ref{ddrate}) over the angle variable
\bea
\frac{d\Gamma^{B\rightarrow K}}{ds}&=&\frac{G_F^2
\alpha^2}{2^{10}\pi^5}|V_{tb}V_{ts}^\ast|^2 m_B^3 \sqrt{\lambda}
\, v \, \Delta \, , \eea where \bea \Delta & = &
\frac{1}{3}~m_B^2~\lambda(3-v^2)(|A_1|^2+|G_1|^2)
+\frac{4m_{\ell}^2}{3s}(12\,r~s+\lambda)|G_1|^2\nnb\\&+&4~m_{\ell}^2~s~|D_1|^2+
s (v^2|E_1|^2+|F_1|^2)+4~m_{\ell}^2 (1-r-s) Re[G_1~D_1^\ast]\nnb\\
&+& 2~m_{\ell}((1-r-s)Re[G_1~F_1^\ast]+s Re[D_1~F_1^\ast]) \,,
\label{deltapi} \eea 
and
\bea \frac{d\Gamma^{B\rightarrow K^\ast}}{ds}&=&\frac{\alpha^2
G_F^2 m_B}{2^{12} \pi^5}|V_{tb} V^*_{ts}|^2
\sqrt{\lambda_{\ast}}~~v~~\Delta_{{\ast}} \eea where \bea
\Delta_{{\ast}} & = & \frac{8}{3}\lambda_{\ast} m_B^6 s
((3-v^2)|A|^2+ 2 v^2 |E|^2)
-\frac{4}{r}\lambda_{\ast} m_B^2 m_{\ell}Re[(F-m_B^2 (1-r_{\ast})G-m_B^2 s H)N^\ast]\nnb \\
&+& \frac{1}{r_{\ast}}\lambda_{\ast} m_B^4 \Bigg [ s v^2
|Q|^2+\frac{1}{3}\lambda_{\ast} m_B^2
(3-v^2)|C|^2+s |N|^2 +m_B^2s^2 (1-v^2)|H|^2 \nnb \\
& + & \frac{2}{3}[(3-v^2)\, W_{\ast}-3 \,s (1-v^2)]
Re[F~G^\ast]-2\, s \,(1-v^2)Re[F~H^\ast] \nnb \\ &+&2 \,m_B^2 s
(1-r_{\ast})(1-v^2)Re[G~H^\ast]+
\frac{2}{3}(3-v^2)W_{\ast} Re[B~C^\ast] \Bigg ]\nnb \\
& + & \frac{1}{3 r_{\ast}} m_B^2 \Bigg [ (\lambda_{\ast} +12
r_{\ast} s)(3-v^2)|B|^2 +\lambda_{\ast} m_B^4 [\lambda_{\ast}
(3-v^2)\nnb \\&-&3 s (s-2 r_{\ast}-2)(1-v^2)]|G|^2 +
(\lambda_{\ast} (3-v^2)+24 r_{\ast} s v^2)|F|^2 \Bigg]
\label{deltarho}. \eea

We would  also like to present the CP asymmetry in the
forward-backward asymmetry $A_{CP}(A_{FB})$, which is another observable that can
give information about the physics beyond the SM.  It is defined as 
\begin{eqnarray}
A_{CP}(A_{FB})& = & \frac{A_{FB} -\bar{A}_{FB}}{A_{FB}
+\bar{A}_{FB}} ~~. \label{ACPAFB1}
\end{eqnarray}
where $\bar{A}_{FB}$ is the CP conjugate of ${A}_{FB}$.

Finally, we would like to discuss the  lepton
polarization effects for the \BKll and \BKsll decays. The
polarization asymmetries of the final lepton is defined as
\begin{eqnarray}
P_{n} (s) & = & \frac{(d\Gamma (S_n)/ds)-(d\Gamma
(-S_n)/ds)}{(d\Gamma (S_n)/ds)+(d\Gamma (-S_n)/ds)} \label{PL}
\end{eqnarray}
for $n=L,~N,~T$. Here, $P_L$, $P_T$ and $P_N$ are the longitudinal,
transversal and normal polarizations, respectively. The unit vectors
$S_n$ are defined as follows:
\bea
S_L&=&(0,\overrightarrow{e}_L)=\Bigg(0,\frac{\overrightarrow{p}_+}{|\overrightarrow{p}_+|}\Bigg)\nnb\\
S_N&=&(0,\overrightarrow{e}_N)=\Bigg(0,\frac{\overrightarrow{p}\times\overrightarrow{p}_+}{|\overrightarrow{p}
\times\overrightarrow{p}_+|}
\Bigg)\nnb\\
S_T&=&(0,\overrightarrow{e}_T)=\Bigg(0,\overrightarrow{e}_N\times
\overrightarrow{e}_L \Bigg)\,,\eea 
where $\overrightarrow{p}=\overrightarrow{p}_{K},\overrightarrow{p}_{K^*}$ and 
$\overrightarrow{p}_+$ are  the three-momenta of $K,K^*$ and $\ell^+$, respectively. 
The longitudinal unit vector
$S_L$ is boosted to the CM frame of $\ell^{+}\ell^{-}$ by Lorentz
transformation: \bea
S_{L,CM}=\Bigg(\frac{|\overrightarrow{p}_+|}{m_\ell},\frac{E_\ell~
\overrightarrow{p}_+}{m_\ell|\overrightarrow{p}_+|}\Bigg).\eea 
It follows from the definition of unit vectors $S_n$ that $P_T$ 
lies in the decay plane while $P_N$ is perpendicular to it, and 
they are not changed by the boost.

After some algebra, we obtain the following expressions 
for the polarization components of the $\ell^+$ lepton in \BKll and \BKsll decays:
\begin{eqnarray}
P_{L}^{B\rightarrow
K}&=&\frac{4m_B^3\,\upsilon}{3\Delta}\Bigg(-2m_B\lambda
Re[A_1~G_1^\ast]-6t(1-r-s)Re[G_1~E_1^\ast]-6st
Re[D_1~E_1^\ast]\nnb\\&+&3s
Re[E_1~F_1^\ast]\Bigg)~,\nnb\\
P_{T}^{B\rightarrow K}&=&\frac{m_B^3
\pi\sqrt{\lambda}}{\sqrt{s}\Delta}\Bigg(-2m_B(1-r-s)t
Re[A_1~G_1^\ast]-2m_Bst
Re[A_1~D_1^\ast]\nnb\\&+&(s-4t^2)Re[G_1~E_1^\ast]+s
Re[A_1~F_1^\ast]\Bigg)~,\nnb\\
P_{N}^{B\rightarrow
K}&=&\frac{m_B^3\pi\upsilon\sqrt{s\lambda}}{\Delta}\Bigg(2m_B t
Im[G_1~D_1^\ast]-Im[A_1~E_1^\ast]-Im[G_1~F_1^\ast]\Bigg)~,
 \eea
and  
 \bea
P_{L}^{B\rightarrow
K^\ast}&=&\frac{4m_B^2\,\upsilon}{r_\ast\Delta_\ast}\Bigg(\lambda_\ast
m_B t
Re[(-F+m_B^2(1-r_\ast)G+m_B^2sH)~Q^\ast]\nnb\\&+&\frac{8}{3}\lambda
m_B^4 r_\ast s Re[A~E^\ast]-\frac{1}{3}Re[B~(\lambda
m_B^2(1-r_\ast-s)G^\ast-(\lambda +12r_\ast s
)F^\ast)\nnb\\&+&C~(\lambda m_B^2(1-r_\ast-s)F^\ast-\lambda^2
m_B^4 G^\ast)]+\frac{1}{2}\lambda_\ast m_B^2 s
Re[Q~N^\ast]\Bigg)~,\nnb\\
P_{T}^{B\rightarrow K^\ast}&=&\frac{\pi m_B^2
\sqrt{\lambda_\ast}}{\Delta_\ast r \sqrt{s}}\Bigg(8m_B^2 r_\ast st
Re[A~B^\ast]-m_B^2
t(1-r_\ast)[(1-r_\ast-s)Re[B~G^\ast]\nnb\\&-&\lambda_\ast m_B^2
Re[C~G^\ast]]-\lambda_\ast m_B^2 t Re[F~C^\ast]-\frac{1}{2}m_B s
Re[(B(1-r_\ast-s)-\lambda_\ast m_B^2
C)N^\ast]\nnb\\&+&\lambda_\ast m_B^4 st Re[C~H^\ast]+\frac{1}{2}s
m_B \upsilon^2 Re[(F(1-r_\ast-s)-\lambda_\ast m_B^2
G)~Q^\ast]\nnb\\&-&t(1-r_\ast-s)Re[(m_B^2s
H-F)B^\ast]\Bigg)~,\nnb\\
P_{N}^{B\rightarrow K^\ast}&=&\frac{\pi m_B^3\upsilon
\sqrt{s\lambda}}{\Delta_\ast r_\ast}\Bigg(4m_B t
r_\ast~Im[B~E^\ast+A~F^\ast]+\frac{1}{2}\lambda_\ast m_B^2
Im[-2m_B t H^\ast G\nnb\\&-&G~N^\ast+C~Q^\ast]-(1+3r_\ast-s)m_B t
Im[G~F^\ast]\nnb\\&-&(1-r_\ast-s)Im[m_Bt
H~F^\ast-\frac{1}{2}N~F^\ast+\frac{1}{2}Q~B^\ast] \Bigg).\nnb \eea


%
\section{Numerical results and discussion \label{s3}}
In this section we present the numerical analysis of the exclusive
decays \BKll and \BKsll in model IV. We will  give the results for only $\ell =\tau$
channel,  which demonstrates the NHB effects more manifestly. The input parameters 
we used in this analysis are as follows:
\begin{eqnarray}
 m_K =0.493 \, GeV \, , \, & & m_B =5.28 \, GeV \, , \, m_b =4.8 \, GeV \, , \,m_c =1.4 \,
GeV \, , \, m_{\tau} =1.77 \, GeV \, ,  \nnb \\ m_{K^*} =0.893 \, GeV \, , \,  & & 
m_{H^{\pm}}=250 \, GeV \, , m_{H^{0}}=125\, GeV\, ,\,m_{h^{0}}=100\, GeV \, \nnb
\\ |V_{tb} V^*_{ts}|=0.04 \, , \,  & &\alpha^{-1}=129 \, ,
\,G_F=1.17 \times 10^{-5}\, GeV^{-2} \,  , \,\tau_B=1.64 \times 10^{-12} \, s \,  .
\end{eqnarray}

The masses of the charged and neutral Higgs bosons, $m_{H^\pm}$, $m_{H^0}$, $m_{A^0}$
and $m_{h^0}$, and  the ratio of the vacuum expectation values
of the two Higgs doublets, $\tan\beta$, remain as free parameters
of the model. The restrictions on $m_{H^\pm}$, and $\tan\beta$ have been
already discussed in section \ref{s1}.  For the masses of the neutral Higgs bosons, 
the lower limits are given as $m_{H^0}\geq 115$ GeV, $m_{h^0}\geq 89.9$ GeV and
$m_{A^0}\geq 90.1$ GeV in \cite{ALEPH}.

Before we present our results, a small note about the calculations of the long-distance
 effects is in order.  There are five possible resonances in the $c\bar{c}$ system that
can contribute to the decays under consideration and to calculate
them, we need to divide the integration region for
$s$ into two parts so that we have 
\bea
4 m^2_{\ell}/m^2_{B} \leq s 
\leq (m_{\psi_2}-0.02)^2/m^2_{B} \,\,\, , \,\,\,
(m_{\psi_2}+0.02)^2/m^2_{B} \leq s \leq (m_B-m_M)^2/m_B^2 \, , \label{limits}
\eea
where $m_{\psi_2}=3.686$ GeV is the mass of the  second resonance, and $M=K,K^*$.

In the following, we give results of our calculations about the dependencies of the
differential branching ratio $dBR/ds$,  forward-backward asymmetry $A_{FB}(s)$, CP violating
asymmetry $A_{CP}(s)$, CP asymmetry in the forward-backward asymmetry $A_{CP}(A_{FB})(s)$
and finally the components of the lepton polarization asymmetries, $P_L(s)$, $P_T(s)$ and 
$P_N(s)$,  of the $B \rightarrow K \tau^+ \tau^-$ and $B \rightarrow K^* \tau^+ \tau^-$ decays
 on the invariant dilepton mass $s$. In order to
investigate the dependencies of the above physical quantities on the model parameters, namely
CP violating phase $\xi$ and $\tan \beta$, we eliminate the other parameter $s$ by performing
the $s$ integrations over the allowed kinematical region (Eq.(\ref{limits})) so as to obtain 
their  averaged values,$<A_{FB}>$, $<A_{CP}>$, $<A_{CP}(A_{FB})>$, $<P_L>$, $<P_T>$ and 
$<P_N>$.

Numerical results are shown in Figs. (\ref{KdBR})-(\ref{PNkcpKs}) and 
we have the following line conventions: dot lines, dashed-dot lines and solid lines
represent the model IV contributions with $\tan \beta =10, 30, 50$, respectively and 
the  dashed lines are for the SM predictions. The cases of switching off NHB contributions
i.e., setting $C_{Q_i}=0$, almost coincide with the cases of 2HDM contributions 
with $\tan \beta =10$, therefore we did not plot them seperately.

In Fig.(\ref{KdBR}), we give  the dependence of the $dBR/ds$ on $s$ for \BKtt. 
From this figure  NHB effects are very obviously seen,  especially in the high-s region.
 
In Fig. (\ref{KdAFB}) and Fig. (\ref{KAFBkcp}),  $A_{FB}(s)$ and $<A_{FB}>$ of \BKtt as a 
function of $s$ and CP violating phase $\xi$ are presented. Since $A_{FB}$
arises in the 2HDM only when NHB effects are taken into account, it provides a good probe
to test these effects.  We see that $A_{FB}$
is quite sensitive to $\tan \beta$ and it is negative for all values  of $\xi$  and s except
in the $\psi^{,}$ region.  $<A_{FB}>$ in \BKtt is between $(-0.04,-0.01)$, which is non zero 
but  hard to observe.  

Fig. (\ref{KdACP}) and Fig. (\ref{KACPkcp}) show the dependence of $A_{CP}(s)$ 
on $s$ and  $<A_{CP}>$  on $\xi$ for \BKtt decay . We see that $A_{CP}(s)$
is quite sensitive to $\tan \beta$ and its sign does not change in the allowed 
values of $s$ except in the resonance mass region of $\psi^{,}$  when $\tan \beta =50$. 
It follows from Fig. (\ref{KACPkcp})
that $<A_{CP}>$ is also sensitive to $\xi$, and it varies in the range
 $(-0.8,0.8)\times 10^{-2}$, which may provide an 
indication for the existence of new physics since  $A_{CP}$ is zero in the SM. 

$A_{CP}(A_{FB})(s)$ and $<A_{CP}(A_{FB})>$ of \BKll as a 
function of $s$ and CP violating phase $\xi$ are presented in Fig. (\ref{KdACPAFB}) and 
Fig. (\ref{KACPAFBkcp}),
respectively. We note that in both of these figures, predictions for the different values of
$\tan \beta$ completely coincide which indicates that  $A_{CP}(A_{FB})$ is  not sensitive to
this parameter in \BKtt decay. As seen from Fig. (\ref{KACPAFBkcp}), $<A_{CP}(A_{FB})>$  
strongly depends on CP violating  phase $\xi$ and it can reach about $6 \%$ for some
values of $\xi$.

In Figs. (\ref{KdPL})-(\ref{KdPN}), we present the $s$  dependence of the 
longitudinal $P_L$, transverse $P_T$ and normal $P_N$ polarizations of the final lepton 
for \BKtt decay. We see  that except the $\psi^{,}$ region, $P_N$ is negative for 
all values of $s$, but $P_L$ and
$P_T$ change sign with the different choices of the values of $\tan \beta$. The effects
of NHB exchanges are also very obvious. 
In Figs. (\ref{KPLkcp}))-(\ref{KPNkcp}),  dependence of the averaged values of the 
longitudinal $<P_L>$, transverse $<P_T>$ and normal $<P_N>$ polarizations of the final 
lepton for \BKll decay  on $\xi$ are shown. It is obvious from these figures that
$<P_L>$ ($<P_N>$) is weakly (strongly) sensitive to $\xi$ while $<P_T>$ is totaly
insensitive to $\xi$. We also note that $<P_N>$ is zero in the SM and it is at the order of
$1\%$ in model IV for $\tan \beta =30$. Thus, measurement of this component in future 
experiments may provide information about the model IV parameters.

Figs. (\ref{dBRKs}) -(\ref{PNkcpKs}) are devoted to the  \BKstt decay. In Fig.(\ref{dBRKs}),  
dependence of the $dBR/ds$ on $s$ is given. 
We see that $dBR/ds$ of this process is not as sensitive to the effects of NHB exchanges as 
\BKtt decay and these effects begin  to be significant when $\tan \beta >40$.

In Fig. (\ref{dAFBKs}) and Fig. (\ref{AFBkcpKs}),  $A_{FB}(s)$ and $<A_{FB}>$ of \BKsll 
as a function of $s$ and CP violating  phase $\xi$ are presented. As in \BKtt decay, 
$A_{FB}$ here  is also quite sensitive to $\tan \beta$, and its magnitude gets smaller than the SM 
prediction with the increasing values of $\tan \beta$. As seen from Fig. (\ref{AFBkcpKs}),
$<A_{FB}>$ in \BKsll is of the order of $10\%$ and   strongly dependent on $\xi$, 
especially when $\tan \beta=50$.

Fig. (\ref{dACPKs}) and Fig. (\ref{ACPkcpKs}) show the dependence of $A_{CP}(s)$ on $s$ and 
$<A_{CP}>$  on $\xi$ for \BKstt decay. We see that $A_{CP}(s)$
is quite sensitive to $\tan \beta$ and $\xi$ and it does not change sign in the allowed 
values of $s$. It follows from Fig. (\ref{ACPkcpKs}) that $<A_{CP}>$ is  of the order 
of $0.1\%$ and hard to observe.

$A_{CP}(A_{FB})(s)$ and $<A_{CP}(A_{FB})>$ of \BKstt as a 
function of $s$ and CP violating  phase $\xi$ are presented in Fig. (\ref{dACPAFBKs}) and 
Fig. (\ref{ACPAFBkcpKs}), respectively. We see that $A_{CP}(A_{FB})$ comes mainly from 
exchanging NHBs and its magnitude  can reach 0.3 exhibiting a strong dependence  on the 
CP-violating phase $\xi$.

In Figs. (\ref{dPLKs})-(\ref{dPNKs}), we present the $s$  dependence of the 
longitudinal $P_L$, transverse $P_T$ and normal $P_N$ polarizations of the final lepton 
for \BKstt decay. 
We see that  NHB exchanges modify the spectrums of $P_T$ and $P_N$ greatly 
while its effect is relatively weak for $P_L$. We also observe  that 
except the $\psi^{,}$ region, $P_L$ is negative for all values of $s$, but $P_T$ and
$P_N$ change sign with the different choices of the values of $\tan \beta$. 
In Figs. (\ref{PLkcpKs})-(\ref{PNkcpKs}),  dependence of the averaged values of the 
longitudinal $<P_L>$, transverse $<P_T>$ and normal $<P_N>$ polarizations of the final 
lepton for \BKsll decay  on $\xi$ are depicted. It is obvious from these figures that
$<P_L>$,  $<P_T>$ and  $<P_N>$ in model IV  are larger as absolute values 
than the corresponding SM predictions. Sensitivity of these observables to the parameter
$\xi$ is significant when $\tan \beta$ is not smaller than 30. 

We now summarize our results:
\begin{itemize}
\item We observe an enhancement in the differential branching ratio for both \BKtt and 
\BKstt processes in model IV compared to the SM when the NHB effects are taken into account.
The NHB effects  are more manifest in \BKtt decay with respect to  \BKstt decay.
\item $A_{FB}$ comes only from NHB contributions in \BKtt, and its average is between
 $(-0.04,-0.01)$, which is non zero but  hard to observe.
However for \BKstt decay,  it is of the order of $10\%$, which should be within the luminosity
reach of coming B factories. 
\item $<A_{CP}>$ is between $(-0.8,0.8)\times 10^{-2}$ and $(-0.3,0.3)\times 10^{-2}$ in
\BKtt and \BKstt decays, respectively.
Since  $A_{CP}$ for these decays is practically zero in the SM, a nonzero value measured 
in future experiments for $A_{CP}$ will be a definite indication of the existence of new
 physics. 
\item  $A_{CP}(A_{FB})$ is at the order of $1 \%$ for \BKtt  decay and 
it is very sensitive to the CP violating phase $\xi$, but not to $\tan \beta$.
As for \BKstt decay, it comes mainly from 
exchanging NHBs, and can be as large as $30 \%$ for some values of $\xi$.

\item Model IV contributions  modify the spectrums of $P_L$, $P_T$  and $P_N$ greatly 
compared to the SM case for both decays. These quantities are sensitive to the NHB effect
and also the CP violating phase $\xi$, except the $P_T$ component for \BKtt decay.
\end{itemize}

Therefore, the experimental investigation of $A_{FB}$, $A_{CP}$, $A_{CP}(A_{FB})$ and
the polarization components in \BKll and \BKsll decays may be quite suitable
for  testing the new physics effects beyond the SM.

\vspace{2cm}
 
{\large{Acknowledgement}}

We would like to thank Shou Hua Zhu for his comments about the previous version of 
this work.

\newpage
\begin{figure}[htb]
\vskip 0truein \centering \epsfxsize=3.8in
\leavevmode\epsffile{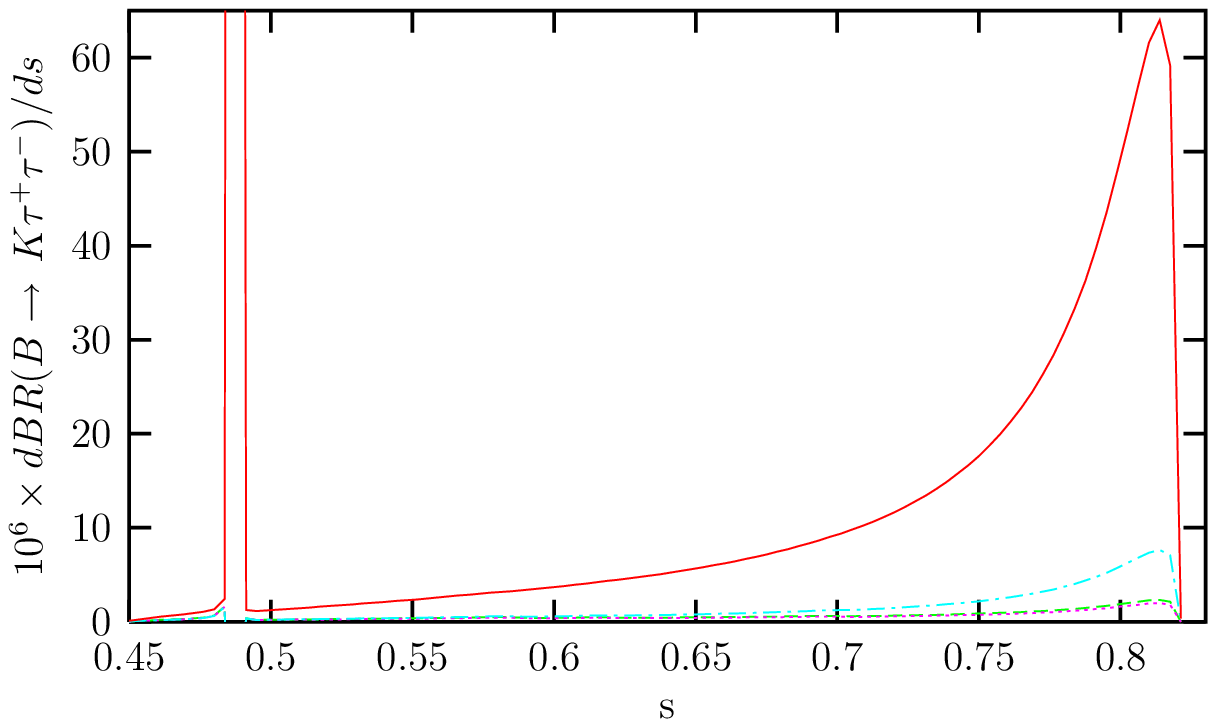} \vskip 0truein \caption[]{The
dependence of the $dBR/ds$ on $s$ for $B\rightarrow K \tau^+ \tau^-$ decay. 
Here dot lines, dashed-dot lines and 
solid lines represent the model IV contributions with $\tan \beta =10, 30, 50$, 
respectively and the  dashed lines are for the SM predictions.} \label{KdBR}
\end{figure}
\begin{figure}[htb]
\vskip 0truein \centering \epsfxsize=3.8in \leavevmode
\epsffile{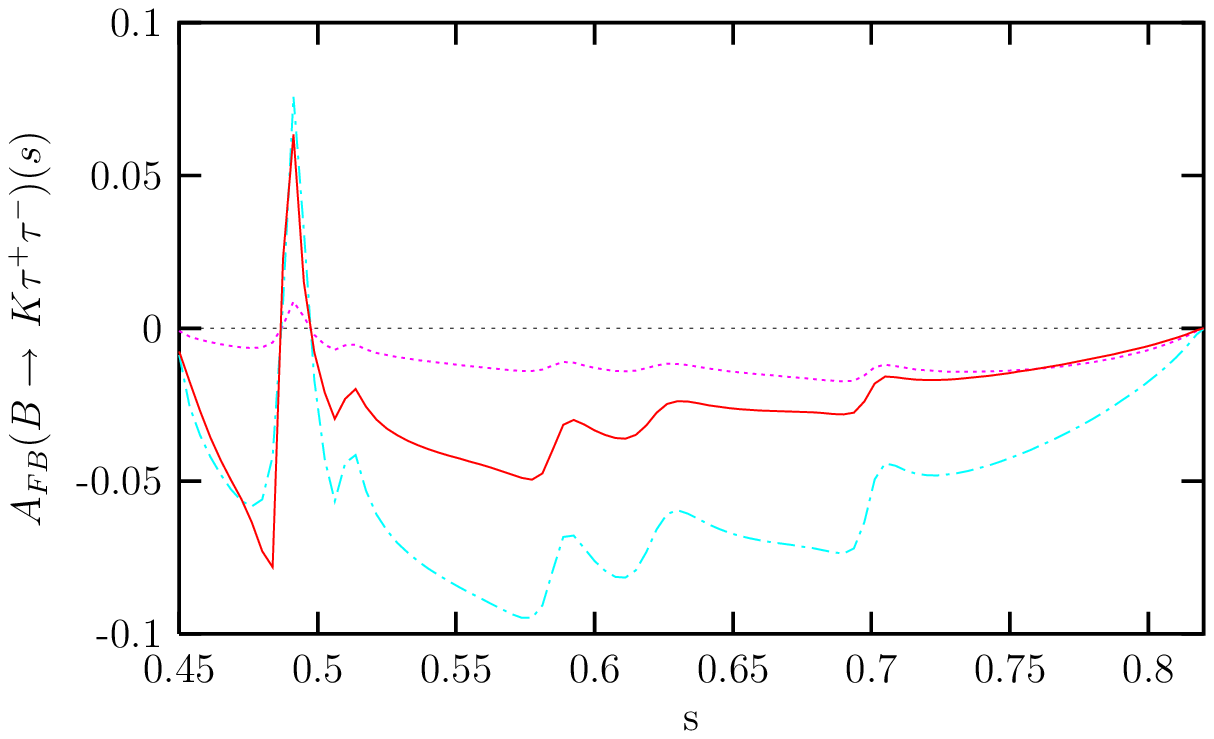} \vskip 0truein \caption{ The dependence
of $A_{FB}(s)(B\rightarrow K \tau^+ \tau^-)$ on  $s$. Here dot lines, dashed-dot lines and 
solid lines represent the model IV contributions with $\tan \beta =10, 30, 50$, 
respectively and the  dashed lines are for the SM predictions.}
\label{KdAFB}
\end{figure}
\begin{figure}[htb]
\vskip 0truein \centering \epsfxsize=3.8in
\leavevmode\epsffile{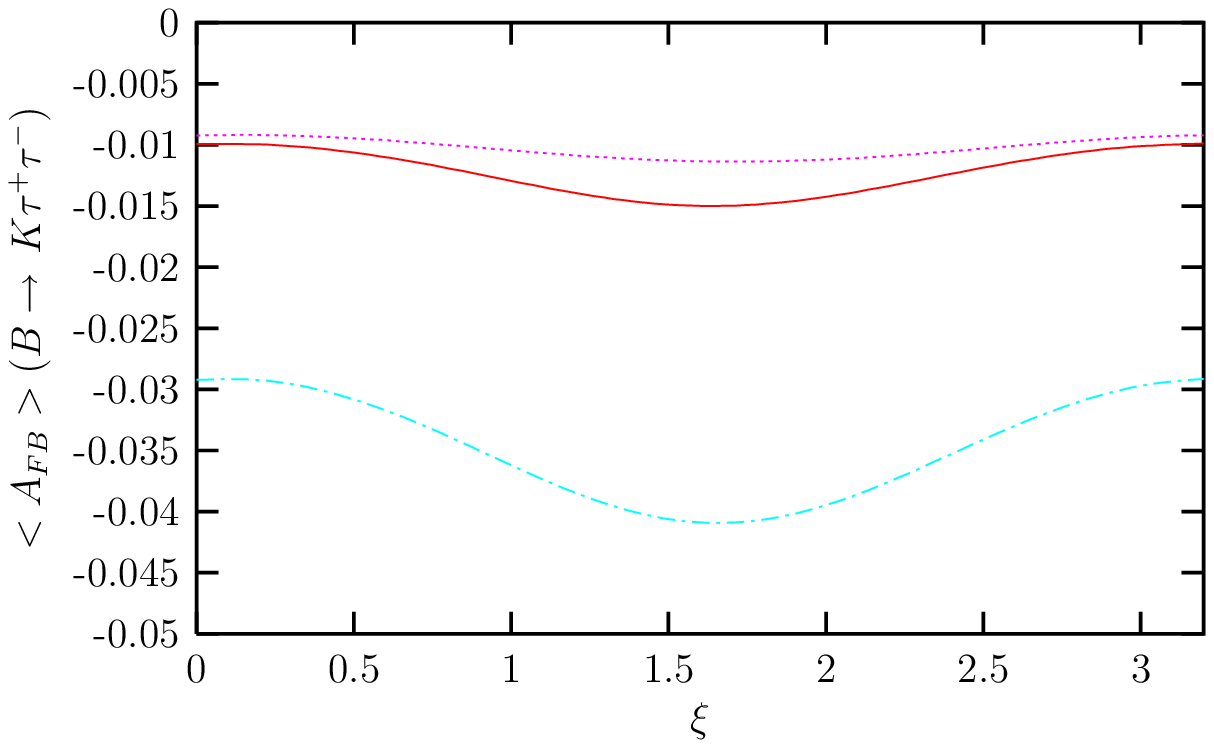} \vskip 0truein \caption{The dependence
of $<A_{FB}>(B\rightarrow K \tau^+ \tau^-)$  on  $\xi$. Here dot lines, dashed-dot lines and 
solid lines represent the model IV contributions with $\tan \beta =10, 30, 50$, 
respectively and the  dashed lines are for the SM predictions.}\label{KAFBkcp}
\end{figure}
\begin{figure}[htb]
\vskip 0truein \centering \epsfxsize=3.8in
\leavevmode\epsffile{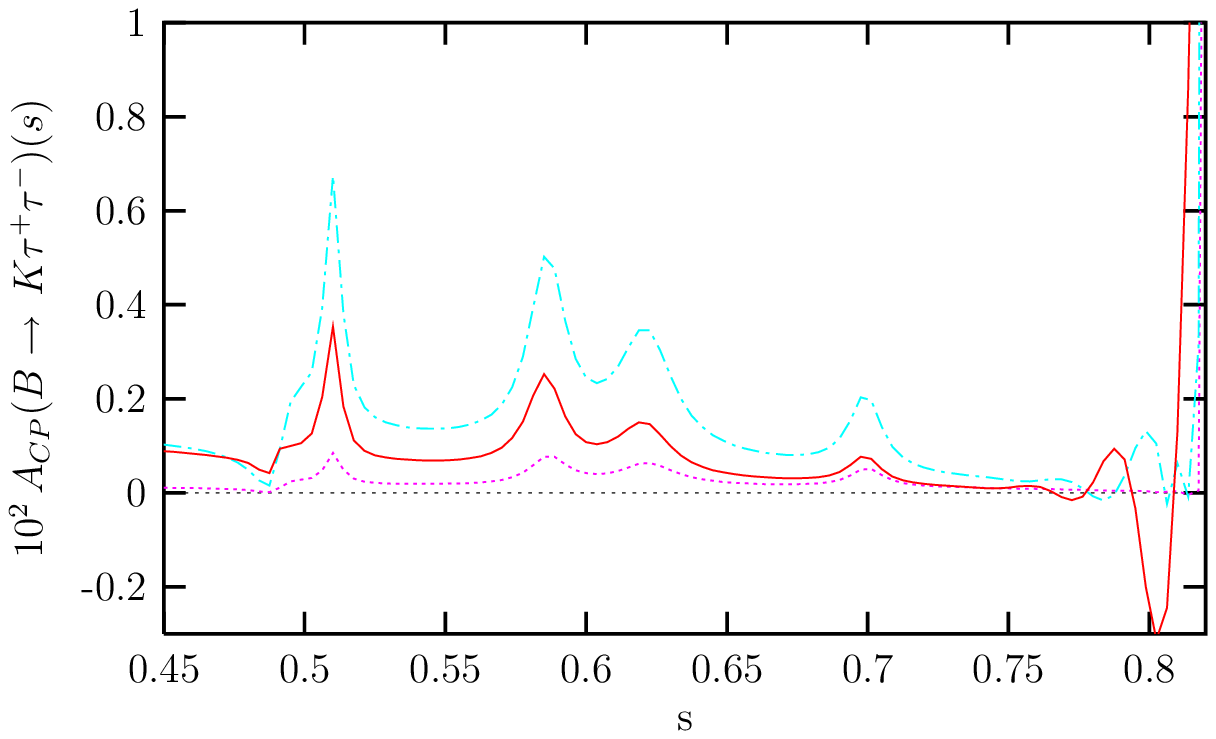} \vskip 0truein \caption{The
same as Fig.(\ref{KdAFB}), but for $A_{CP}(s)(B\rightarrow K \tau^+ \tau^-)$.}
\label{KdACP}
\end{figure}
\begin{figure}[htb]
\vskip 0truein \centering \epsfxsize=3.8in
\leavevmode\epsffile{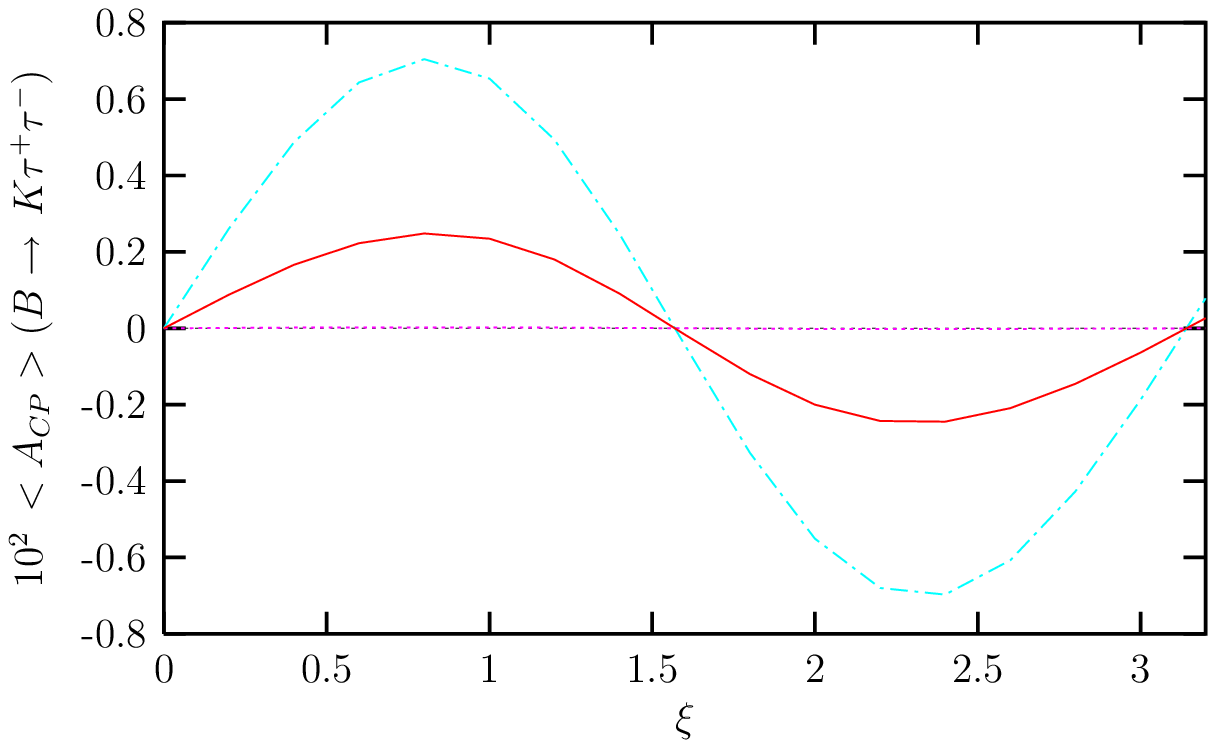} \vskip 0truein \caption{The
same as Fig.(\ref{KAFBkcp}), but for $<A_{CP}> (B\rightarrow K \tau^+ \tau^-)$.}
 \label{KACPkcp}
\end{figure}
\begin{figure}[htb]
\vskip 0truein \centering \epsfxsize=3.8in
\leavevmode\epsffile{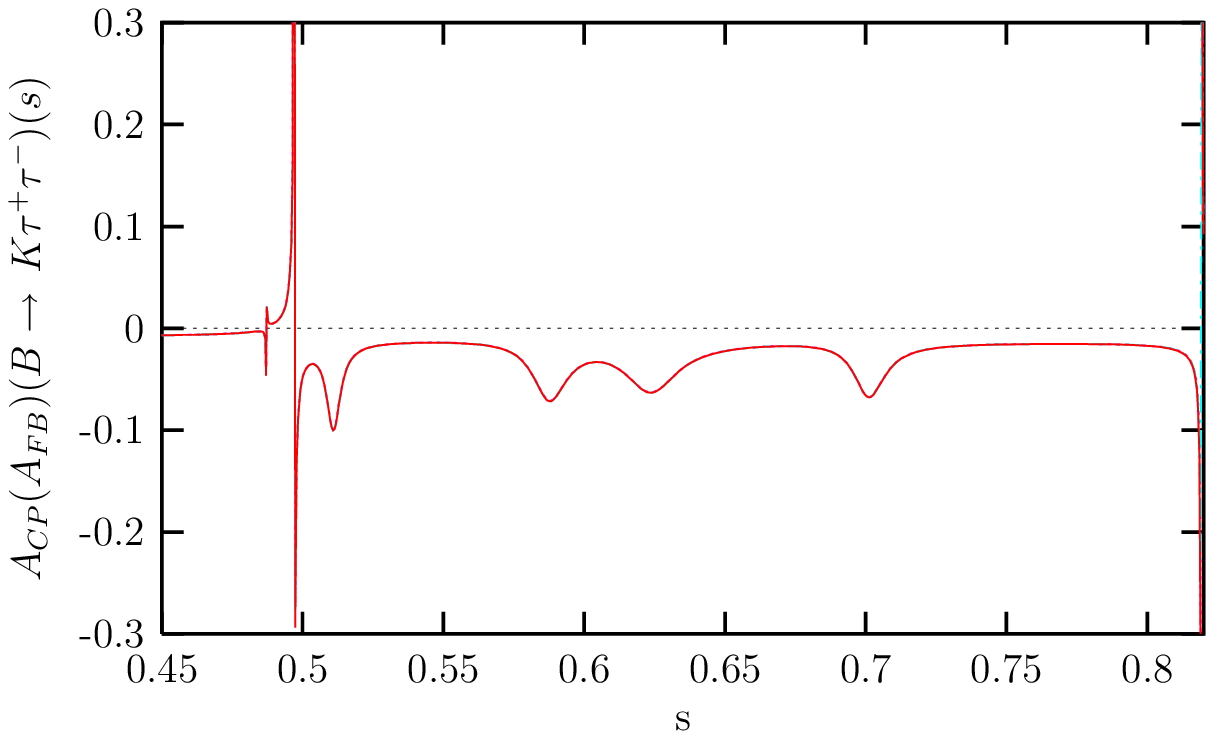} \vskip 0truein \caption{The
same as Fig.(\ref{KdAFB}), but for $A_{CP}(A_{FB})(s)(B\rightarrow K \tau^+ \tau^-)$.} 
\label{KdACPAFB}
\end{figure}
\begin{figure}[htb]
\vskip 0truein \centering \epsfxsize=3.8in
\leavevmode\epsffile{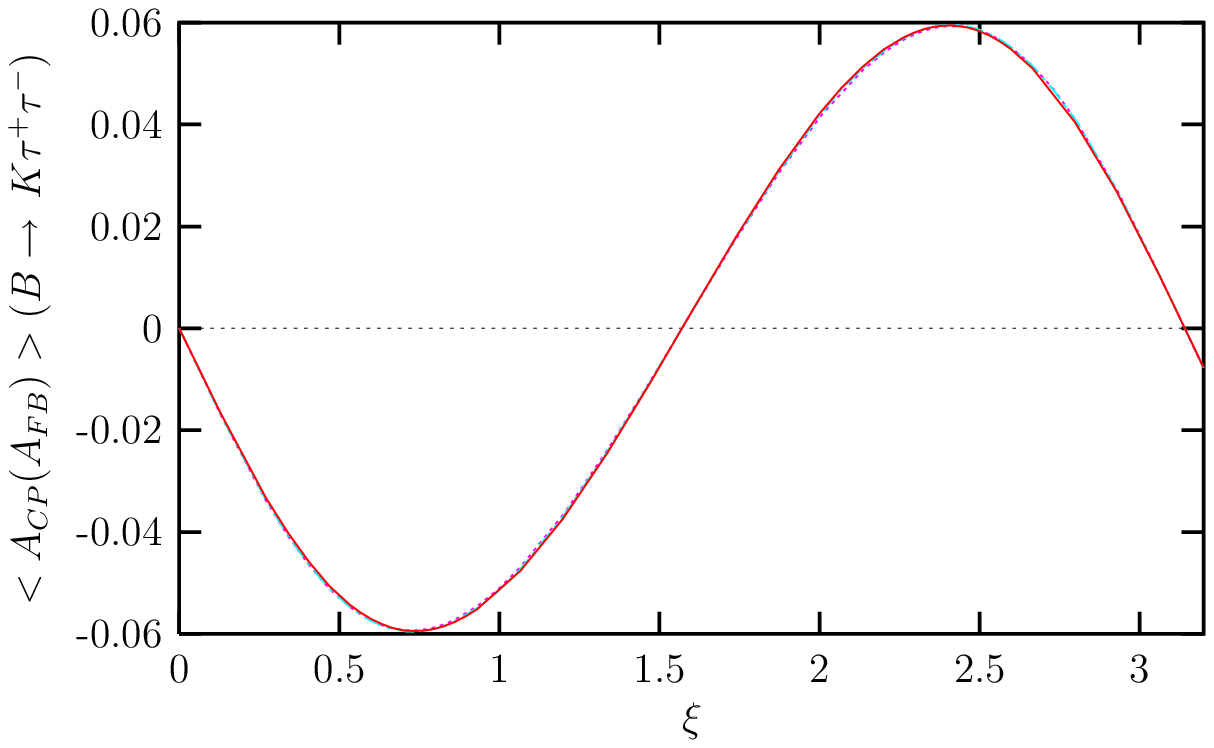} \vskip 0truein \caption{ The
same as Fig.(\ref{KAFBkcp}), but for $<A_{CP}(A_{FB})>(B\rightarrow K \tau^+ \tau^-)$.}
\label{KACPAFBkcp}
\end{figure}
\begin{figure}[htb]
\vskip 0truein \centering \epsfxsize=3.8in
\leavevmode\epsffile{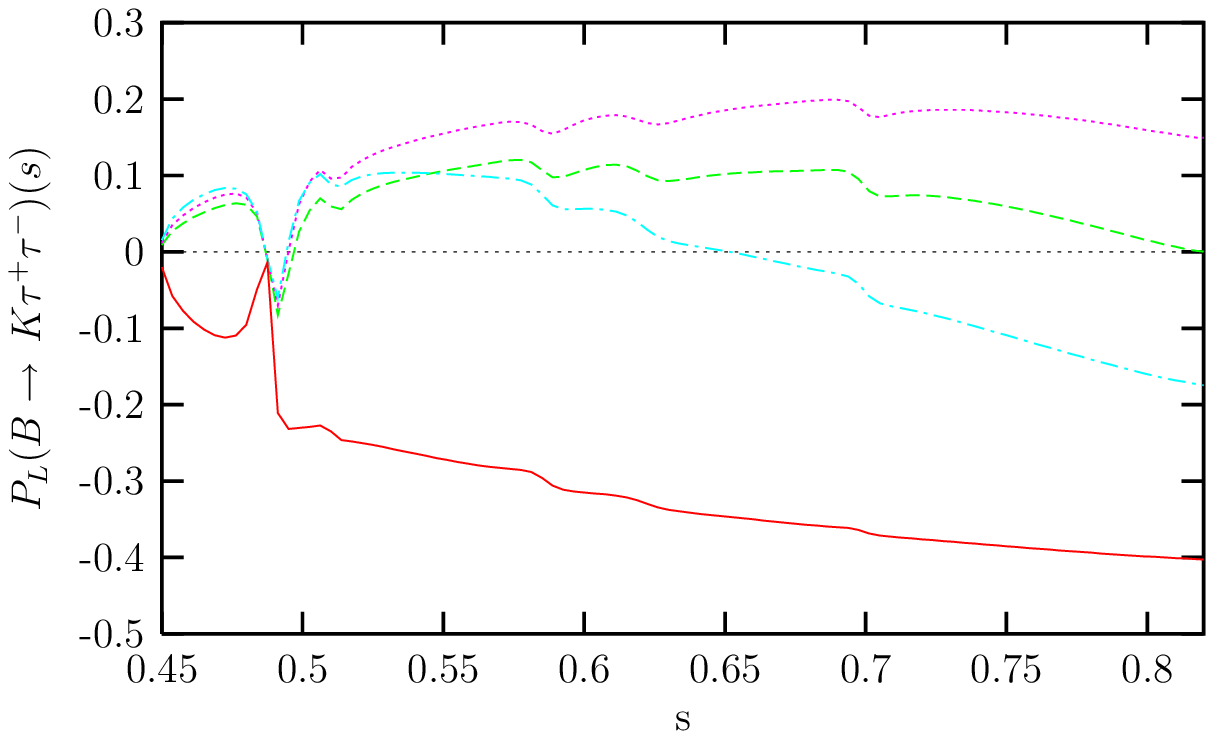} \vskip 0truein \caption{The dependence
of $P_L (s)(B\rightarrow K \tau^+ \tau^-)$ on  $s$. Here dot lines, dashed-dot lines and 
solid lines represent the model IV contributions with $\tan \beta =10, 30, 50$, 
respectively and the  dashed lines are for the SM predictions.}
\label{KdPL}
\end{figure}
\begin{figure}[htb]
\vskip 0truein \centering \epsfxsize=3.8in
\leavevmode\epsffile{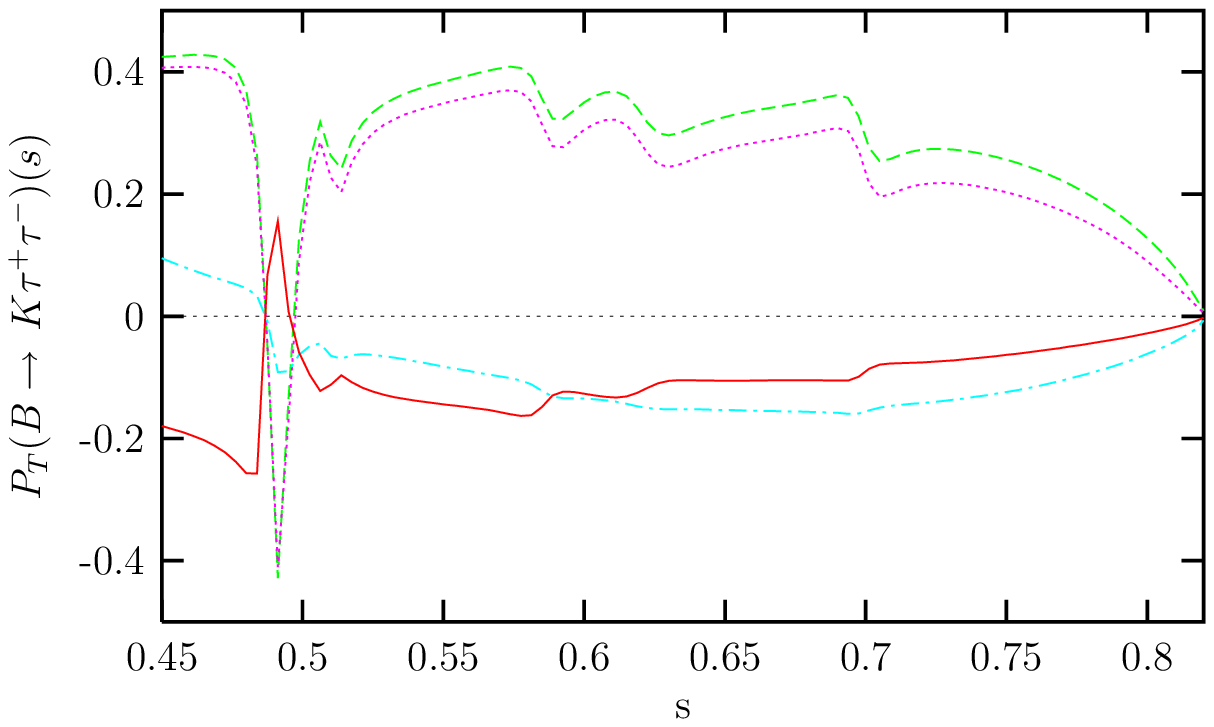} \vskip 0truein \caption{The
same as Fig.(\ref{KdPL}), but for $P_T (s)(B\rightarrow K \tau^+ \tau^-)$.}
\label{KdPT}
\end{figure}
\begin{figure}[htb]
\vskip 0truein \centering \epsfxsize=3.8in
\leavevmode\epsffile{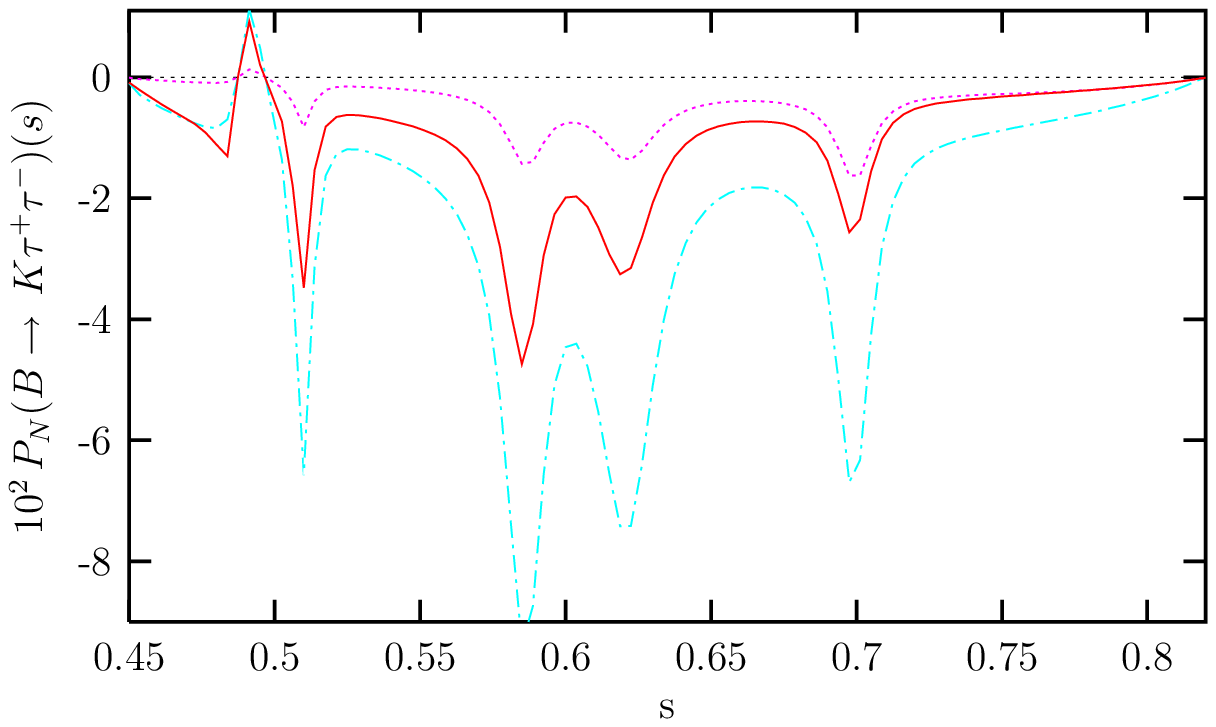} \vskip 0truein \caption{The
same as Fig.(\ref{KdPL}), but for $P_N (s)(B\rightarrow K \tau^+ \tau^-)$.}
\label{KdPN}
\end{figure}
\begin{figure}[htb]
\vskip 0truein \centering \epsfxsize=3.8in
\leavevmode\epsffile{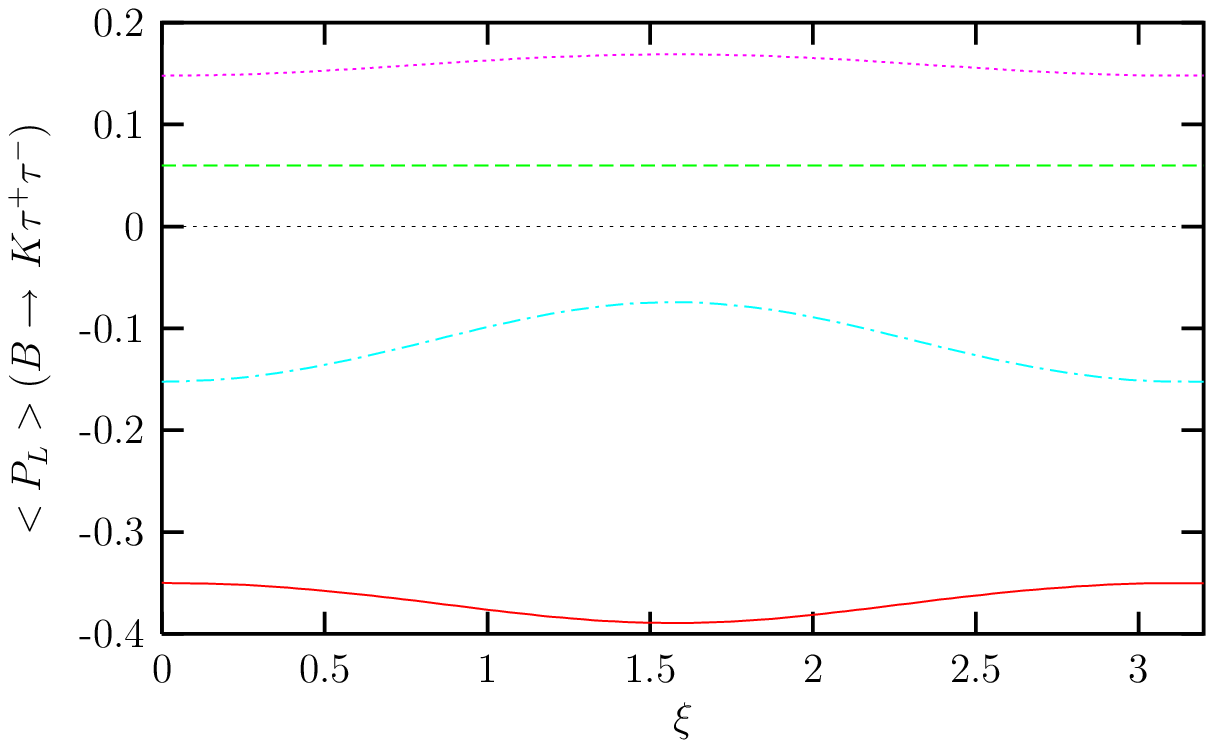} \vskip 0truein \caption{The dependence
of $<P_L >(B\rightarrow K \tau^+ \tau^-)$ on  $\xi$. Here dot lines, dashed-dot lines and 
solid lines represent the model IV contributions with $\tan \beta =10, 30, 50$, 
respectively and the  dashed lines are for the SM predictions.}
\label{KPLkcp}
\end{figure}
\begin{figure}[htb]
\vskip 0truein \centering \epsfxsize=3.8in
\leavevmode\epsffile{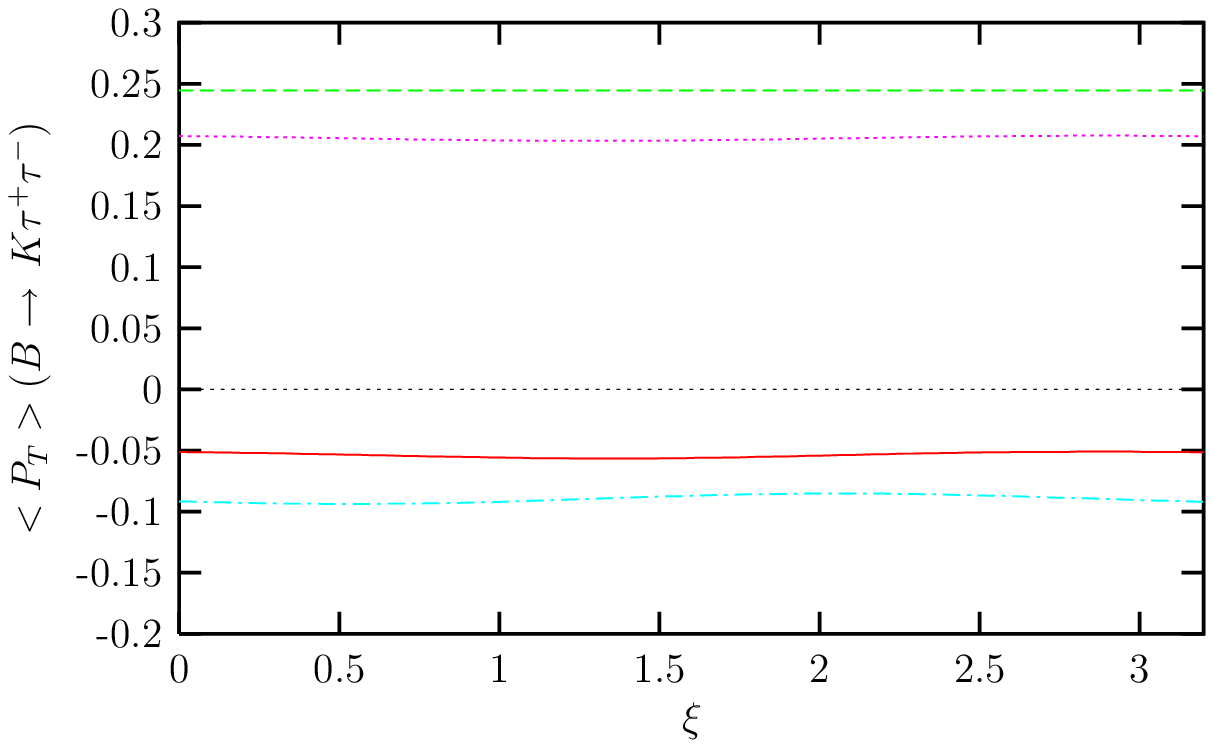} \vskip 0truein \caption{The
same as Fig.(\ref{KPLkcp}), but for $<P_T>(B\rightarrow K \tau^+ \tau^-)$.}
\label{KPTkcp}
\end{figure}
\begin{figure}[htb]
\vskip 0truein \centering \epsfxsize=3.8in
\leavevmode\epsffile{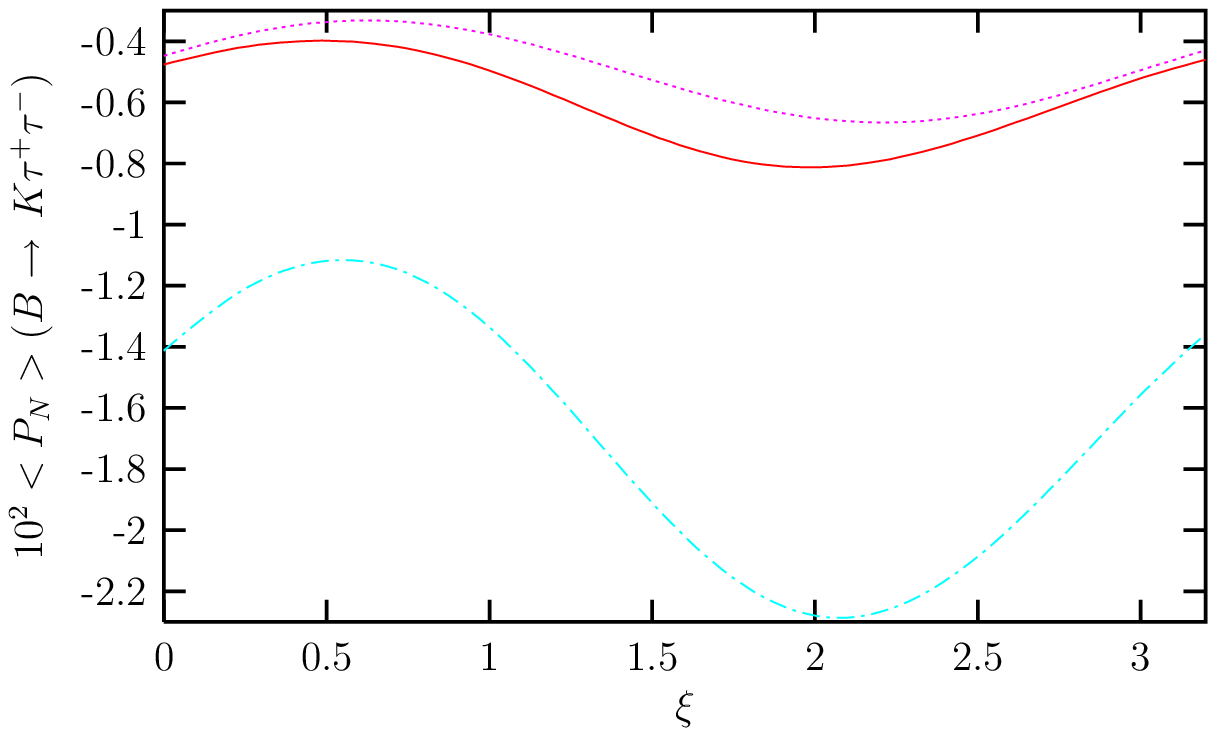} \vskip 0truein \caption{The
same as Fig.(\ref{KPLkcp}), but for $<P_N>(B\rightarrow K \tau^+ \tau^-)$.}
\label{KPNkcp}
\end{figure}
\begin{figure}[htb]
\vskip 0truein \centering \epsfxsize=3.8in
\leavevmode\epsffile{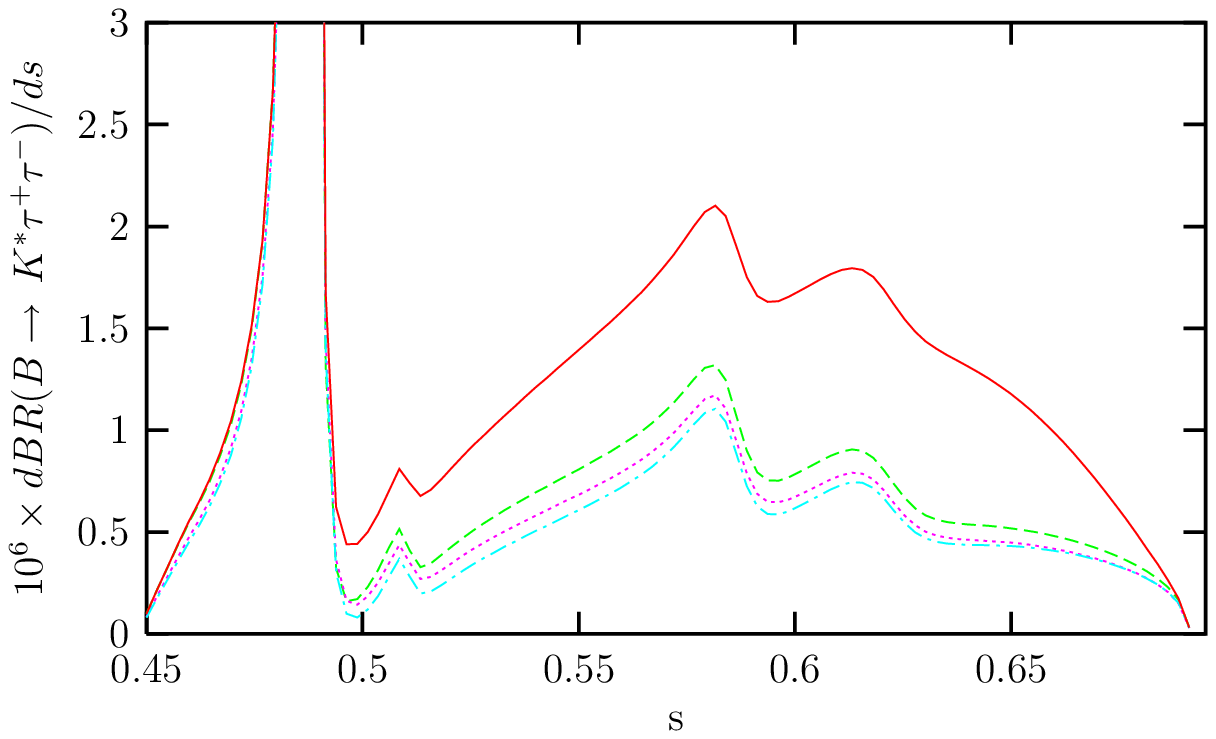} \vskip 0truein \caption[]{The
dependence of the $dBR/ds$ on $s$ for $B\rightarrow K^* \tau^+ \tau^-$ decay. 
Here dot lines, dashed-dot lines and 
solid lines represent the model IV contributions with $\tan \beta =10, 30, 50$, 
respectively and the  dashed lines are for the SM predictions.} \label{dBRKs}
\end{figure}
\begin{figure}[htb]
\vskip 0truein \centering \epsfxsize=3.8in \leavevmode
\epsffile{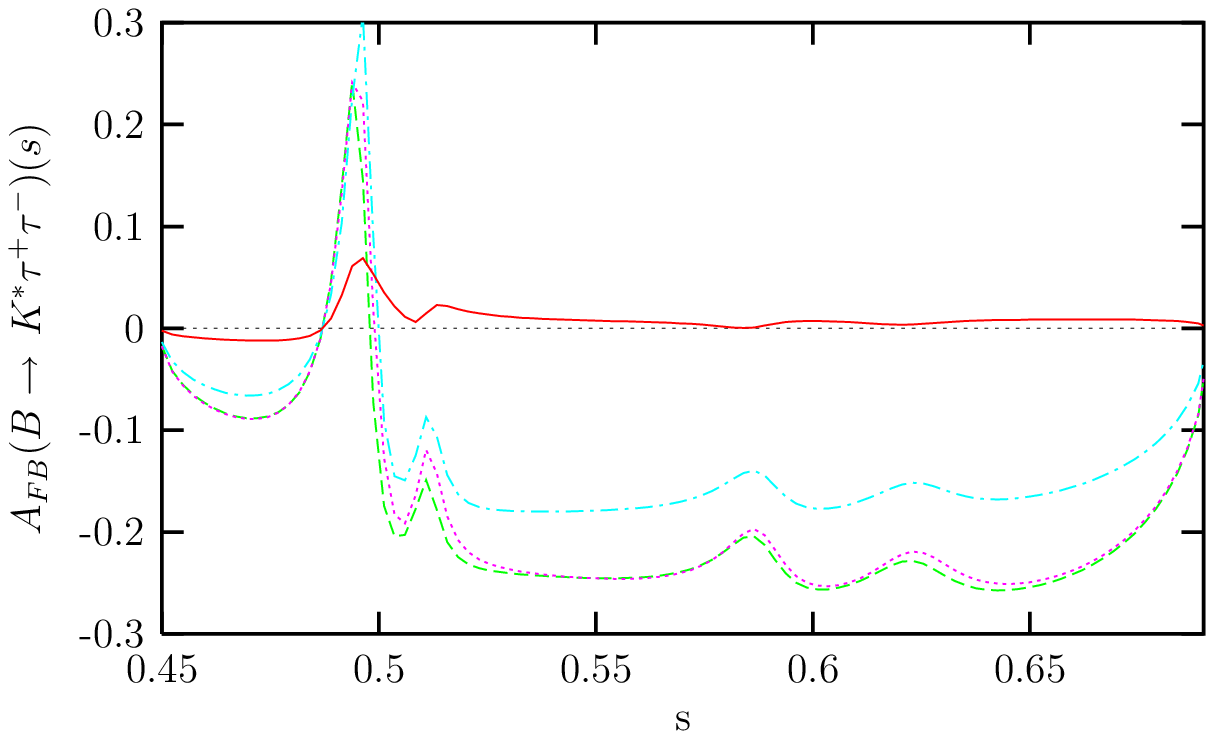} \vskip 0truein \caption{ The dependence
of $A_{FB}(s)(B\rightarrow K^* \tau^+ \tau^-)$ on  $s$. 
Here dot lines, dashed-dot lines and 
solid lines represent the model IV contributions with $\tan \beta =10, 30, 50$, 
respectively and the  dashed lines are for the SM predictions.}
\label{dAFBKs}
\end{figure}
\begin{figure}[htb]
\vskip 0truein \centering \epsfxsize=3.8in
\leavevmode\epsffile{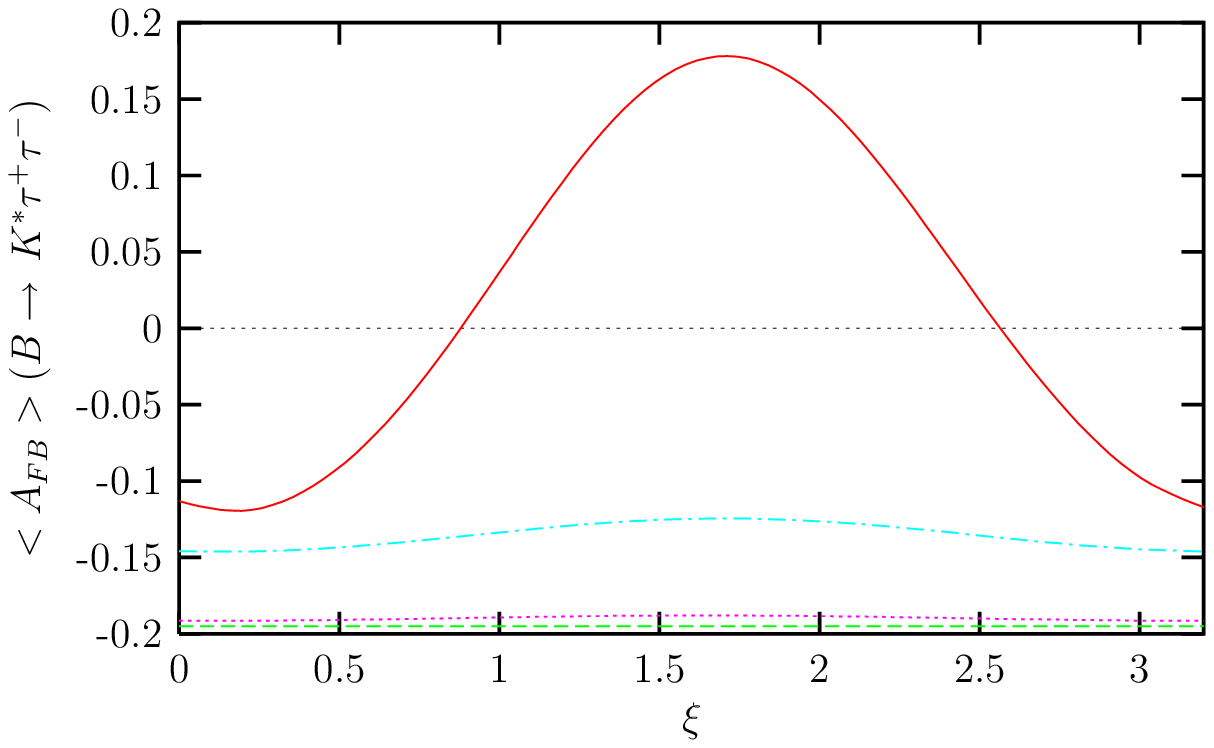} \vskip 0truein \caption{The dependence
of $<A_{FB}>(B\rightarrow K^* \tau^+ \tau^-)$  on  $\xi$. Here dot lines, dashed-dot 
lines and 
solid lines represent the model IV contributions with $\tan \beta =10, 30, 50$, 
respectively and the  dashed lines are for the SM predictions.}\label{AFBkcpKs}
\end{figure}
\newpage
\begin{figure}[htb]
\vskip 0truein \centering \epsfxsize=3.8in
\leavevmode\epsffile{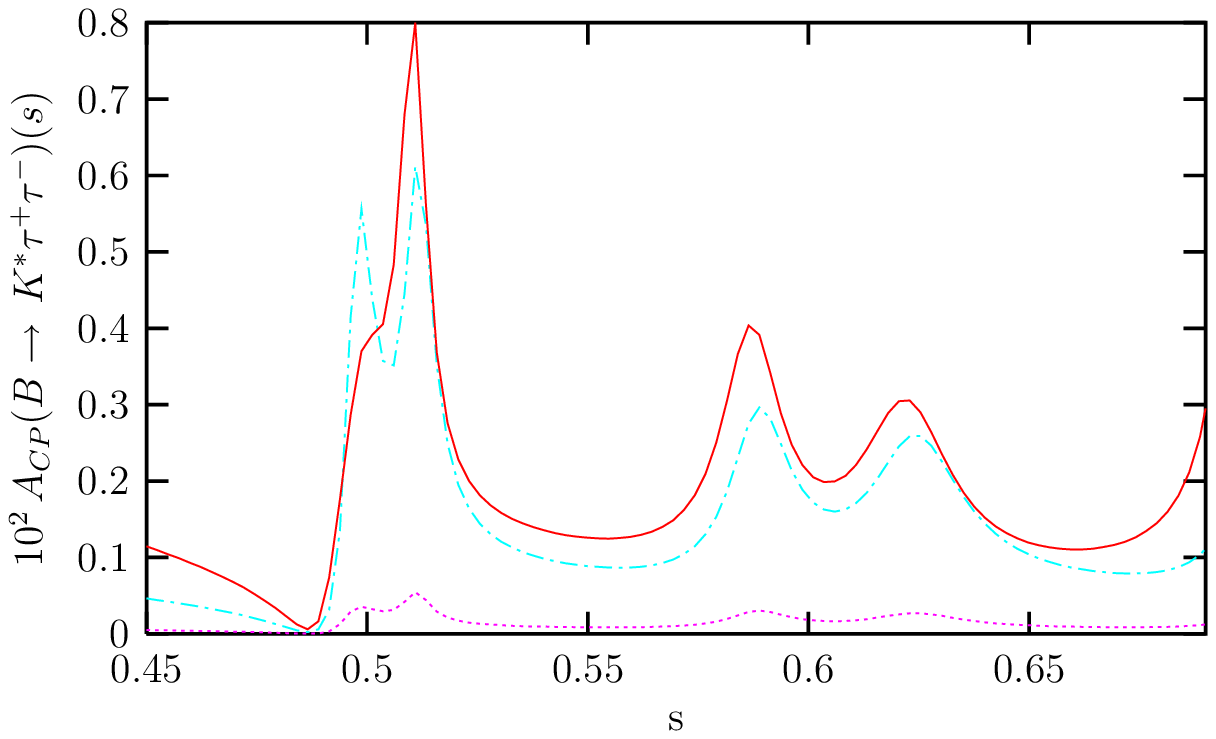} \vskip 0truein \caption{The
same as Fig.(\ref{dAFBKs}), but for $A_{CP}(s)(B\rightarrow K^* \tau^+ \tau^-)$.}
\label{dACPKs}
\end{figure}
\begin{figure}[htb]
\vskip 0truein \centering \epsfxsize=3.8in
\leavevmode\epsffile{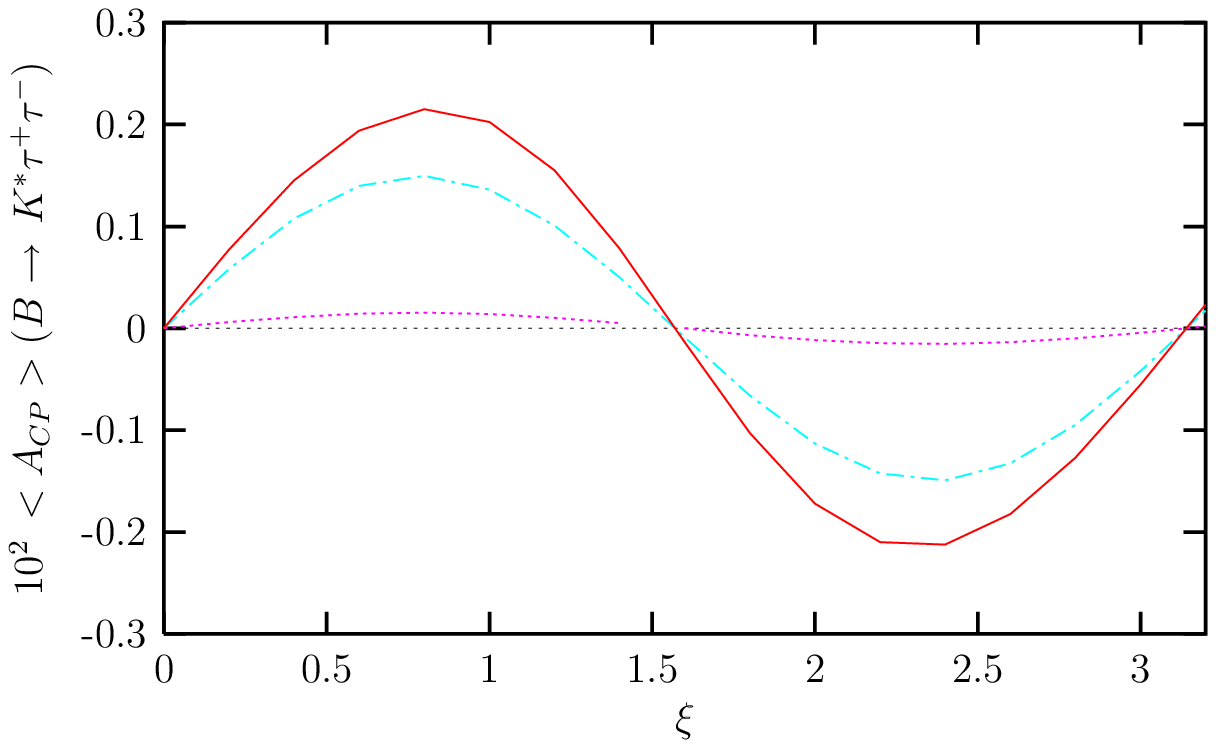} \vskip 0truein \caption{The
same as Fig.(\ref{AFBkcpKs}), but for $<A_{CP}> (B\rightarrow K^* \tau^+ \tau^-)$.}
 \label{ACPkcpKs}
\end{figure}
\newpage
\begin{figure}[htb]
\vskip 0truein \centering \epsfxsize=3.8in
\leavevmode\epsffile{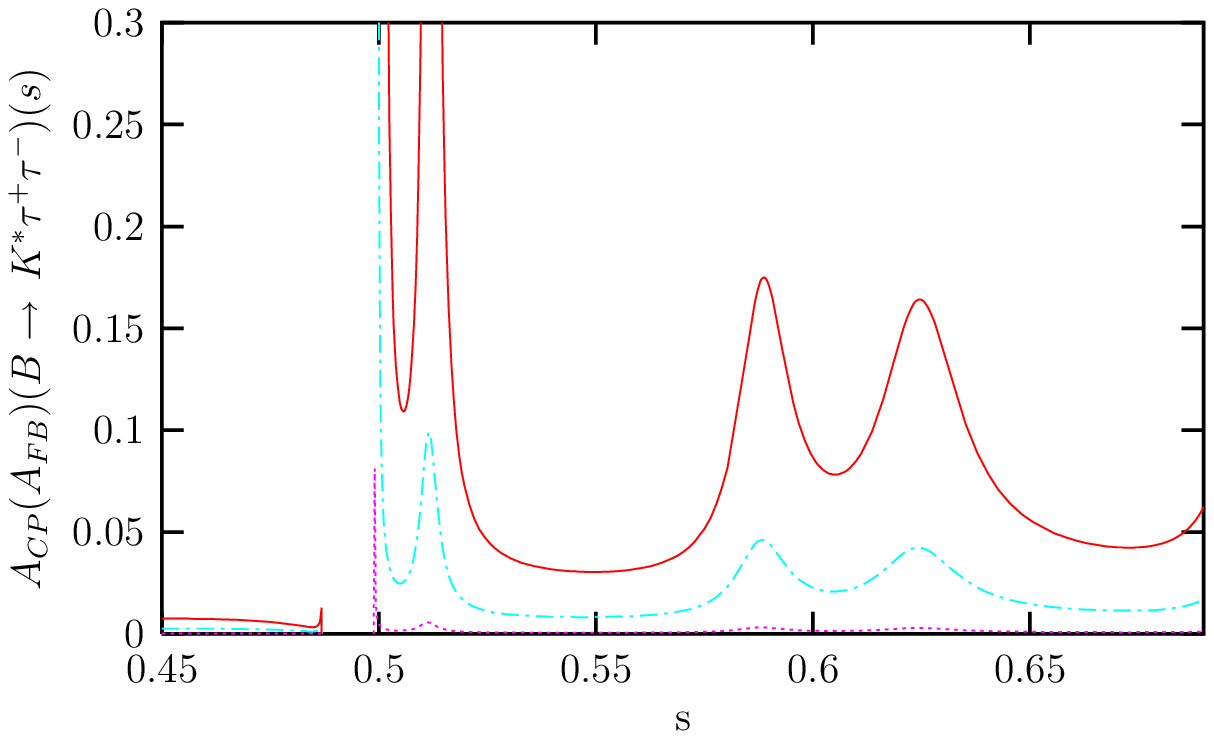} \vskip 0truein \caption{The
same as Fig.(\ref{dAFBKs}), but for $A_{CP}(A_{FB})(s)(B\rightarrow K^* \tau^+ \tau^-)$.} 
\label{dACPAFBKs}
\end{figure}
\begin{figure}[htb]
\vskip 0truein \centering \epsfxsize=3.8in
\leavevmode\epsffile{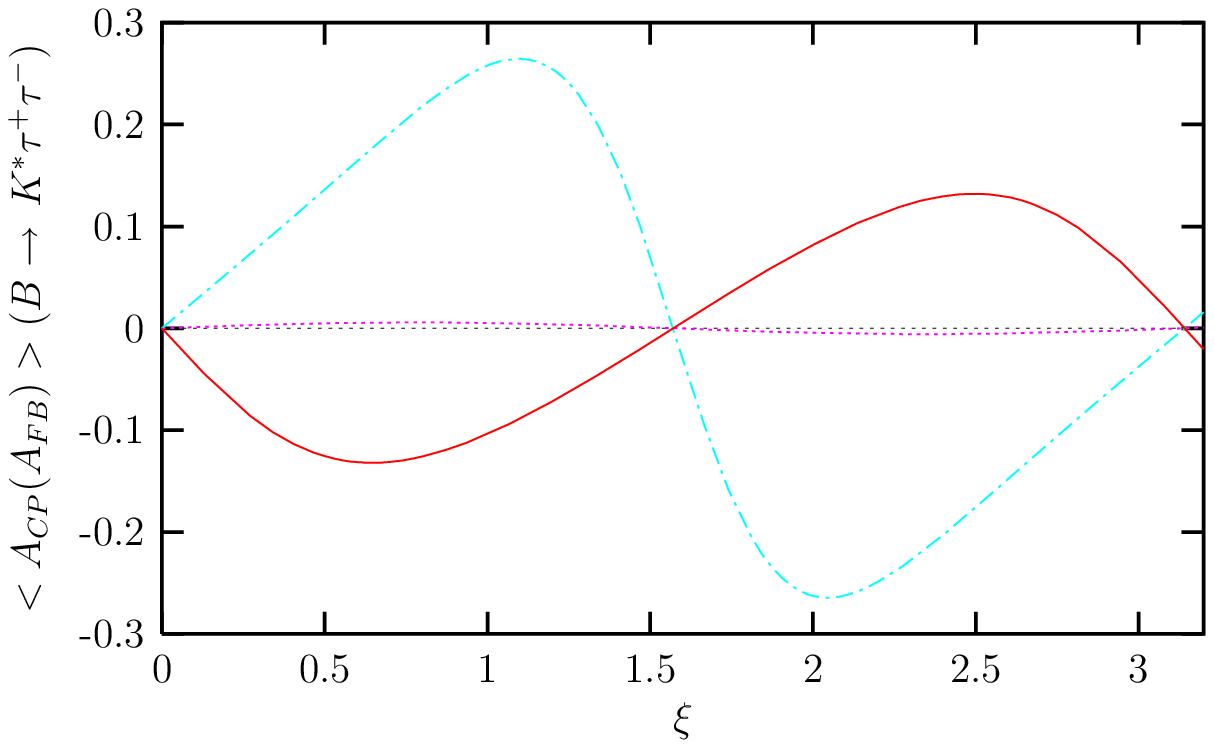} \vskip 0truein \caption{ The
same as Fig.(\ref{AFBkcpKs}), but for $<A_{CP}(A_{FB})>(B\rightarrow K^* \tau^+ \tau^-)$.}
\label{ACPAFBkcpKs}
\end{figure}
\clearpage
\begin{figure}[tbh]
\vskip 0truein \centering \epsfxsize=3.5in
\leavevmode\epsffile{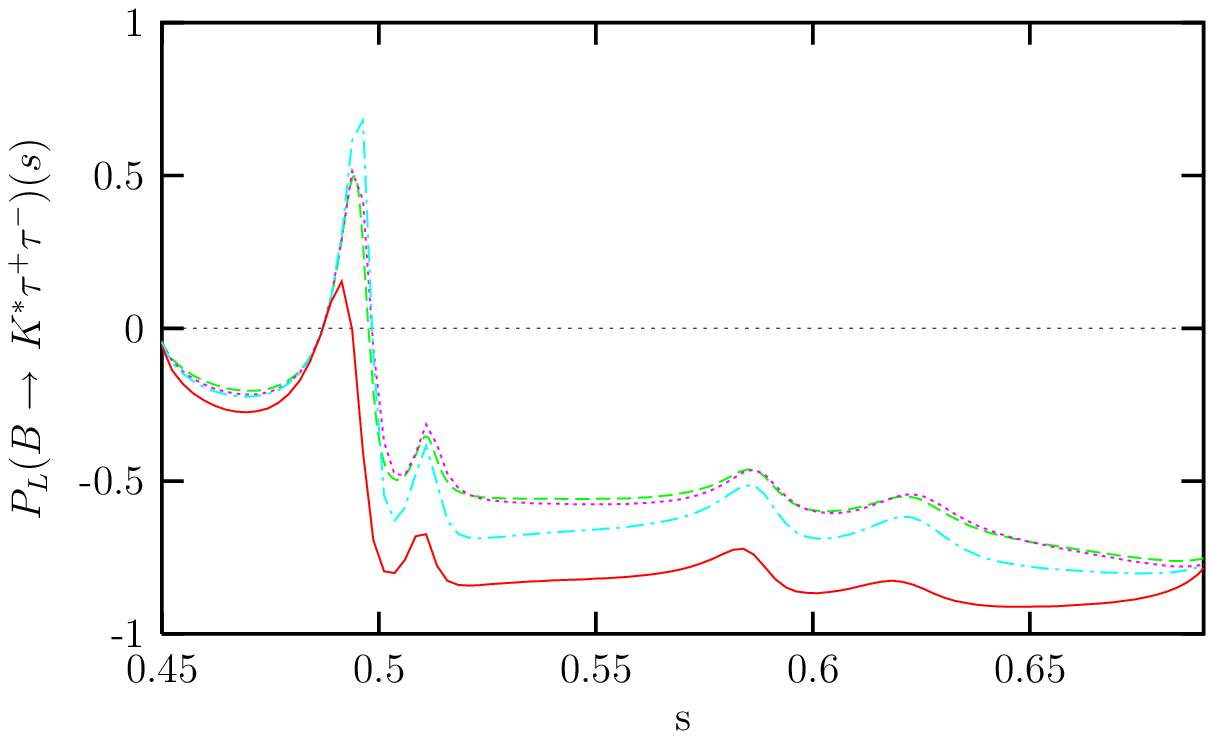} \vskip 0truein \caption{The dependence
of $P_L (s)(B\rightarrow K^* \tau^+ \tau^-)$ on  $s$. Here dot lines, dashed-dot lines and 
solid lines represent the model IV contributions with $\tan \beta =10, 30, 50$, 
respectively and the  dashed lines are for the SM predictions.}
\label{dPLKs}
\end{figure}
\begin{figure}[htb]
\vskip 0truein \centering \epsfxsize=3.5in
\leavevmode\epsffile{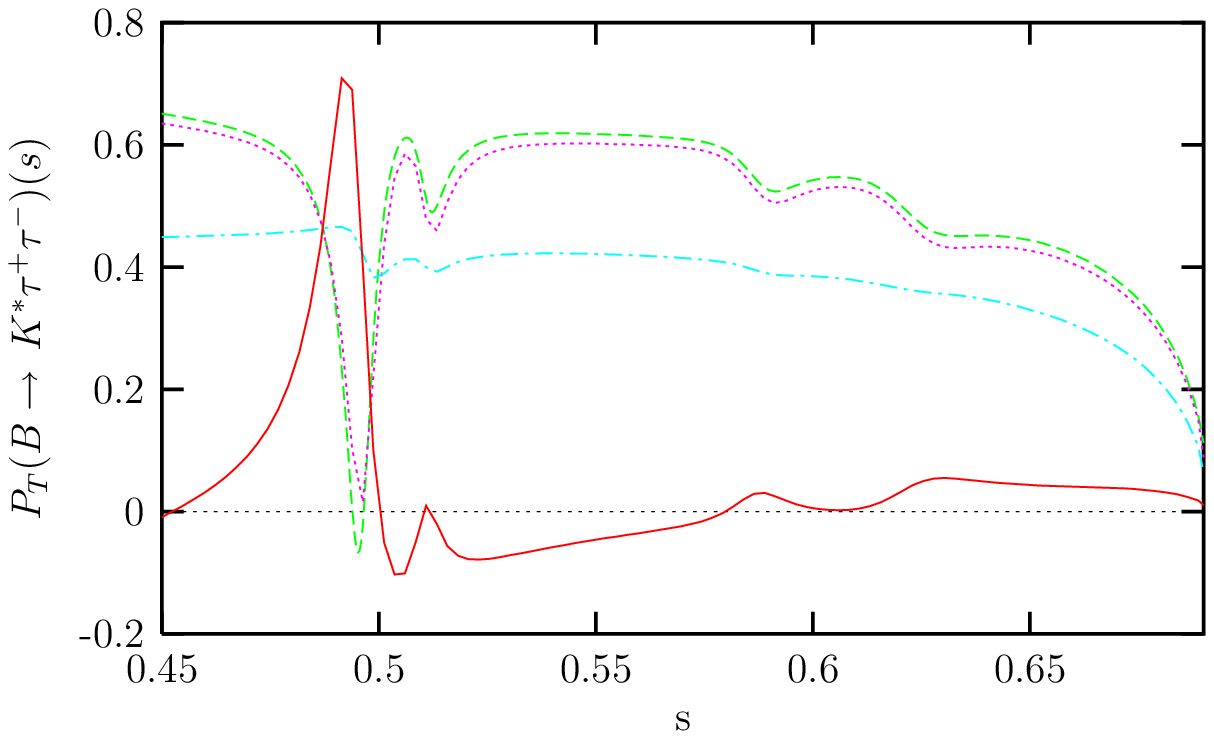} \vskip 0truein \caption{The
same as Fig.(\ref{dPLKs}), but for $P_T (s)(B\rightarrow K^* \tau^+ \tau^-)$.}
\label{dPTKs}
\end{figure}
\begin{figure}[htb]
\vskip 0truein \centering \epsfxsize=3.5in
\leavevmode\epsffile{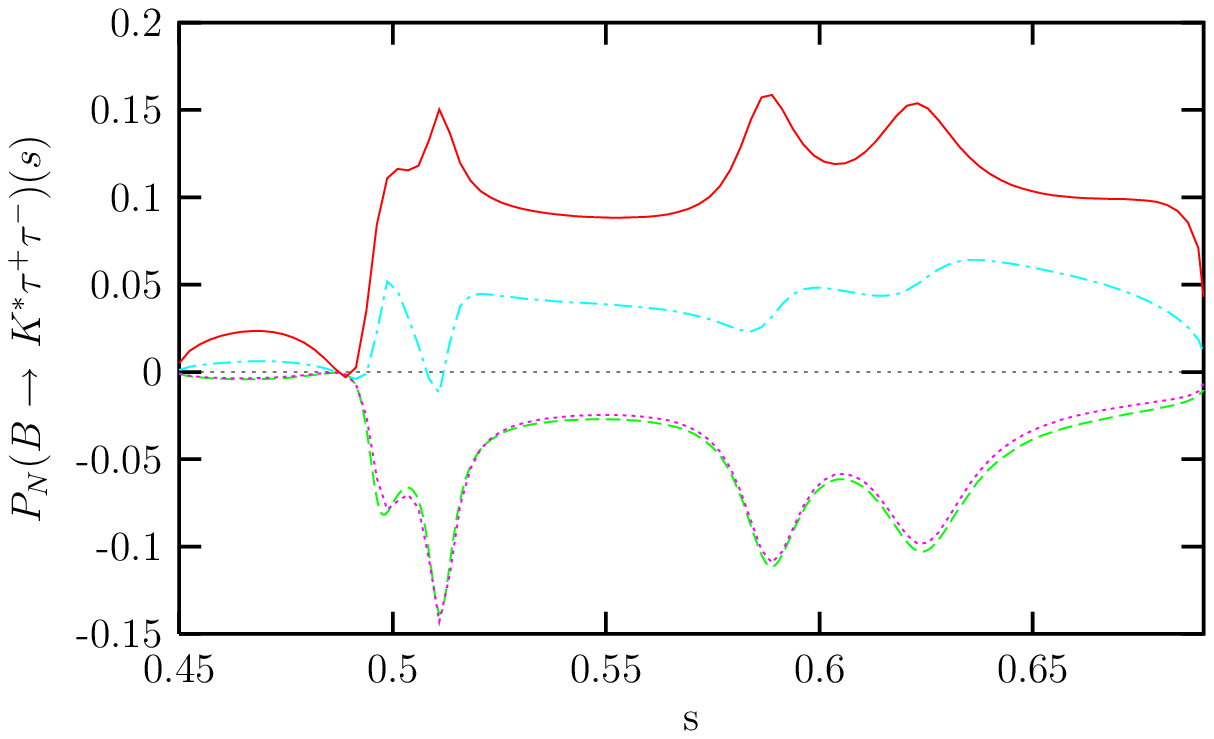} \vskip 0truein \caption{The
same as Fig.(\ref{dPLKs}), but for $P_N (s)(B\rightarrow K^* \tau^+ \tau^-)$.}
\label{dPNKs}
\end{figure}
\begin{figure}[htb]
\vskip 0truein \centering \epsfxsize=3.5in
\leavevmode\epsffile{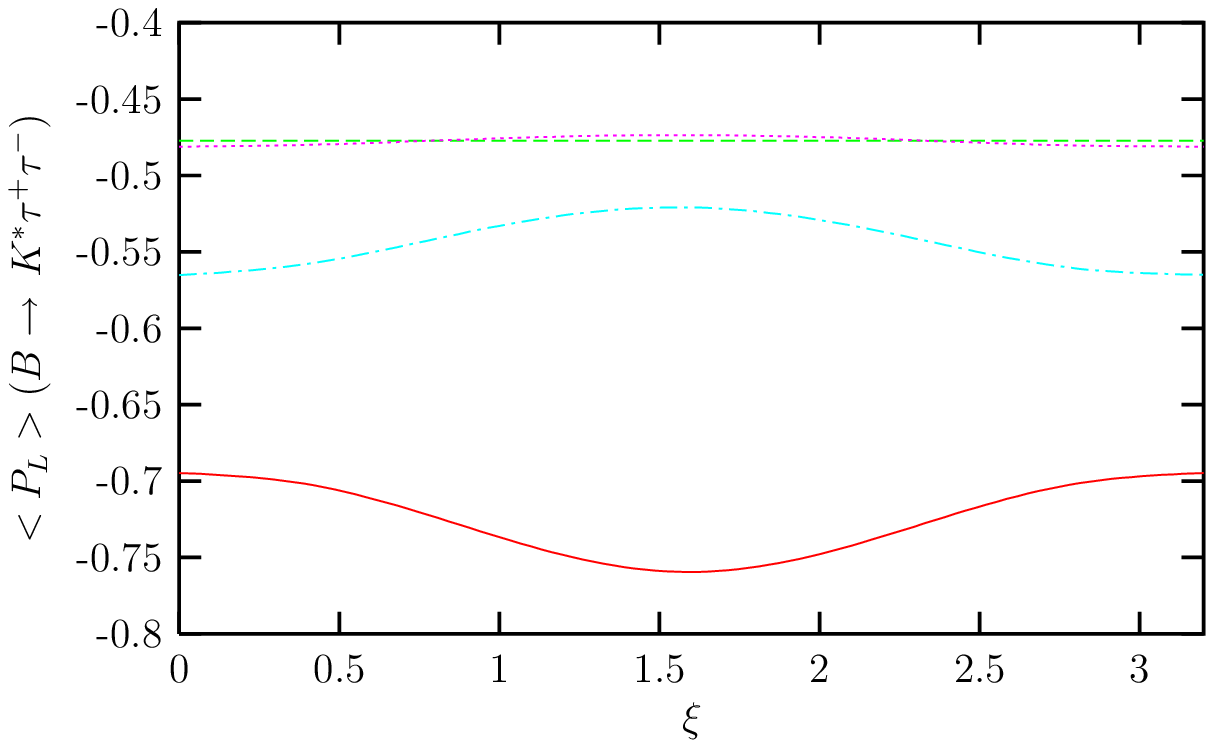} \vskip 0truein \caption{The dependence
of $<P_L >(B\rightarrow K^* \tau^+ \tau^-)$ on  $\xi$. Here dot lines, dashed-dot lines and 
solid lines represent the model IV contributions with $\tan \beta =10, 30, 50$, 
respectively and the  dashed lines are for the SM predictions.}
\label{PLkcpKs}
\end{figure}
\begin{figure}[htb]
\vskip 0truein \centering \epsfxsize=3.5in
\leavevmode\epsffile{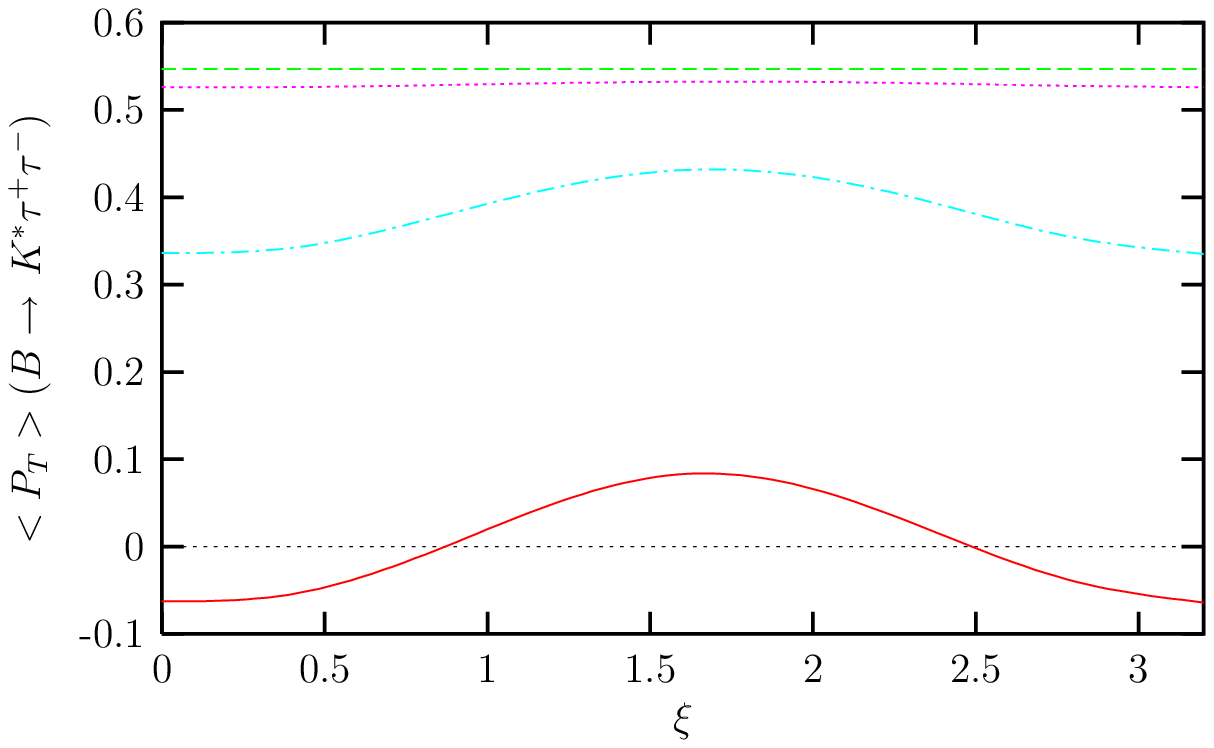} \vskip 0truein \caption{The
same as Fig.(\ref{PLkcpKs}), but for $<P_T>(B\rightarrow K^* \tau^+ \tau^-)$.}
\label{PTkcpKs}
\end{figure}
\begin{figure}[htb]
\vskip 0truein \centering \epsfxsize=3.5in
\leavevmode\epsffile{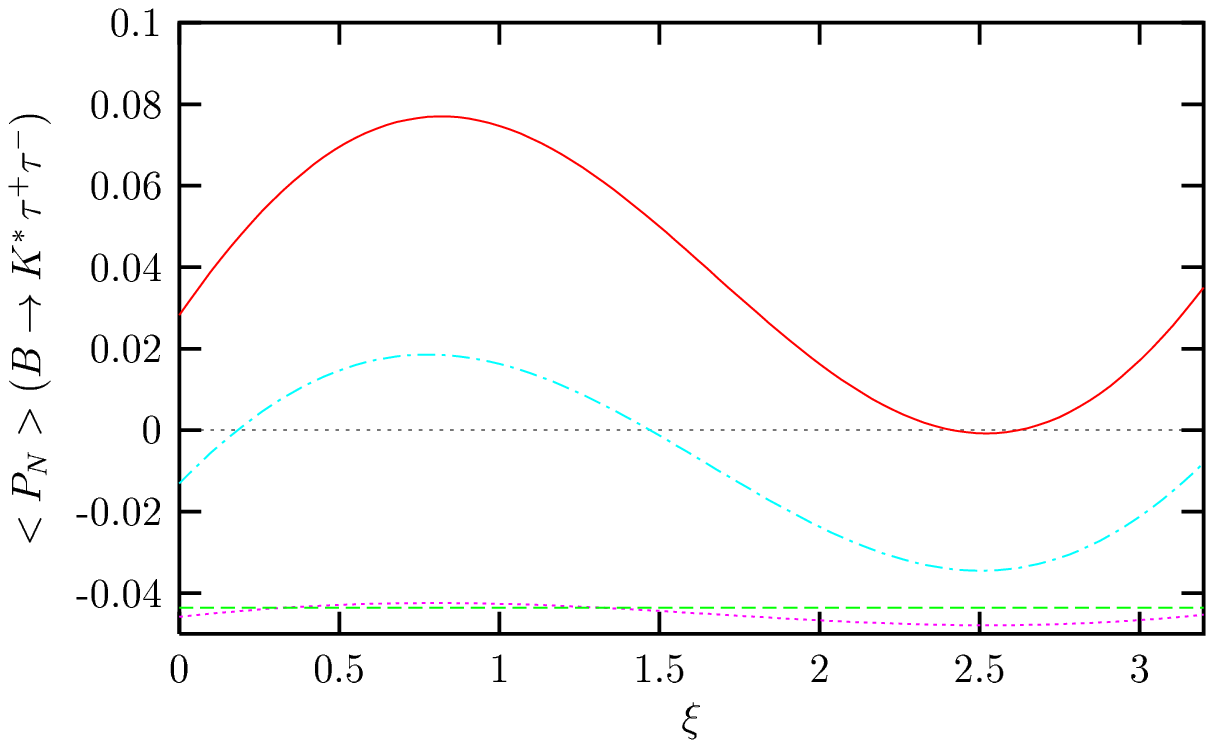} \vskip 0truein \caption{The
same as Fig.(\ref{PLkcpKs}), but for $<P_N>(B\rightarrow K^* \tau^+ \tau^-)$.}
\label{PNkcpKs}
\end{figure}

\end{document}